\documentclass[namedreferences]{solarphysics}
\usepackage[hyperref,optionalrh,solaromanenum]{spr-sola-addons} 
\usepackage{graphicx}                    
\usepackage{color}                       
\usepackage{breakurl}					
\usepackage{todonotes}                   
\usepackage{caption}

\begin{document}

\begin{article}

\begin{opening}

\title{Determining the Intrinsic CME Flux Rope Type Using Remote-sensing Solar Disk Observations}

%
\author[addressref={aff1},corref,email={erika.palmerio@helsinki.fi}]{\inits{E. Palmerio}\fnm{E. Palmerio}\lnm{}}
\author[addressref={aff1},corref,email={}]{\inits{E. Kilpua}\fnm{E. K. J. Kilpua}\lnm{}}
\author[addressref={aff2},corref,email={}]{\inits{A. W. James}\fnm{A. W. James}\lnm{}}
\author[addressref={aff2},corref,email={}]{\inits{L. Green}\fnm{L. M. Green}\lnm{}}
\author[addressref={aff1},corref,email={}]{\inits{J. Pomoell}\fnm{J. Pomoell}\lnm{}}
\author[addressref={aff1},corref,email={}]{\inits{A. Isavnin}\fnm{A. Isavnin}\lnm{}}
\author[addressref={aff2},corref,email={}]{\inits{G. Valori}\fnm{G. Valori}\lnm{}}

%
\runningauthor{Palmerio et al.}
\runningtitle{Magnetic structure of flux ropes}

\address[id={aff1}]{University of Helsinki, Department of Physics, P.O. Box 64, 00014 Helsinki, Finland}
\address[id={aff2}]{University College London, Mullard Space Science Laboratory, Holmbury St. Mary, Dorking, Surrey, RH5 6NT, UK}

\begin{abstract}
A key aim in space weather research is to be able to use remote-sensing observations of the solar atmosphere to extend the lead time
of predicting the geoeffectiveness of a coronal mass ejection (CME). In order to achieve this, the magnetic structure of the CME as it leaves the Sun must be known. In this article we address this issue by developing a method to determine the intrinsic flux rope type of a CME solely from solar disk observations. We use several well known proxies for the magnetic helicity sign, the axis orientation, and the axial magnetic field direction to predict the magnetic structure of the interplanetary flux rope. We present two case studies: the 2 June 2011 and the 14 June 2012 CMEs. Both of these events erupted from an active region and, despite having clear \textit{in situ} counterparts, their eruption characteristics were relatively complex. The first event was associated with an active region filament that erupted in two stages, while for the other event the eruption originated from a relatively high coronal altitude and the source region did not feature the presence of a filament. Our magnetic helicity sign proxies include the analysis of magnetic tongues, soft X-ray and/or extreme-ultraviolet sigmoids, coronal arcade skew, filament emission and absorption threads, and filament rotation.  
Since the inclination of the post-eruption arcades was not clear, we use the tilt of the polarity inversion line to determine the flux rope axis orientation, and coronal dimmings to determine the flux rope footpoints and, therefore, the direction of the axial magnetic field. The comparison of the estimated intrinsic flux rope structure to \textit{in situ} observations at the Lagrangian point L1 indicated a good agreement with the predictions. Our results highlight the flux rope type determination techniques that are particularly useful for active region eruptions, where most geoeffective CMEs originate. 
\end{abstract}

%

\end{opening}

%
\section{Introduction}
\label{s:introduction}

One of the most prominent manifestations of solar activity are coronal mass ejections (CMEs). These are large eruptions of plasma and magnetic field that are expelled from the Sun and propagate into the heliosphere. CMEs form in the solar atmosphere and all models predict that their magnetic field configuration when they leave the lower corona is that of a twisted bundle of magnetic field known as a flux rope \citep[\textit{e.g.},][]{antiochos1999,moore2001,kliem2006}.
However, when CMEs are detected in interplanetary space, they present a diverse range of configurations and signatures \citep[\textit{e.g.},][]{gosling1990,richardson2004a,zurbuchen2006}, and only about one third of them present a well-defined flux rope structure \citep[\textit{e.g.},][]{gosling1990,richardson2004b,huttunen2005}. This is likely to be due to either
a large crossing distance from the flux rope center \citep[\textit{e.g.},][]{cane1997,jian2006,kilpua2011}, deformation of the magnetic field due to interactions between multiple CMEs \citep{burlaga2002}, and/or a significant erosion of the initial magnetic flux \citep{dasso2007,ruffenach2012}, which means that a coherent flux rope is either hard to identify or no longer present. When a flux rope is present, it can be identified in interplanetary space when \textit{in situ} data show a monotonic rotation of the magnetic field direction through a large angle, a low plasma temperature, and a low plasma $\beta$ \citep{burlaga1981}.

Interplanetary CMEs (or ICMEs) with flux ropes are key drivers of intense magnetic storms \citep[\textit{e.g.},][]{gosling1991,webb2000,huttunen2005}. The most important parameter that determines a CME geoeffectiveness is the interplanetary magnetic field (IMF), in particular the magnitude and duration of its north\,--\,south component ($B_Z$) in the Geocentric solar magnetospheric (GSM) coordinate system. The general field pattern of a CME can be determined in terms of the ``flux rope type'' \citep{bothmer1998,mulligan1998}. The type is determined by the direction of the flux rope axial magnetic field, the orientation (tilt) of its axis with respect to the ecliptic plane, and its magnetic chirality. Chirality (or magnetic helicity sign) is the sense of twist of the flux rope (either right-handed or left-handed twist).
Flux ropes that have their axis closely aligned with the ecliptic plane will exhibit a change of sign in $B_Z$ as the ICME passes over the spacecraft. These CMES are known as ``bipolar'' or low inclination clouds. Whereas flux ropes that have their axis orientated perpendicular to the ecliptic plane will maintain the sign of $B_Z$ and are therefore known as ``unipolar'' or high inclination clouds.
For example, in a bipolar ``north-east-south''-type cloud (NES) the field rotates in a right-handed sense (\textit{i.e.} with positive chirality) from north to south, being eastward at the center.

Knowledge of the intrinsic magnetic structure of an erupting CME is vital if long-term space weather forecasting is to be realized. Knowing the flux rope type at the Sun, and therefore the profile of $B_Z$ through the structure, could provide this information a few days in advance of the CME reaching the first Lagrange point (L1), where \textit{in situ} magnetic field measurements are typically made.
However, due to rotations, deflections, and deformations the actual field structure that impacts the Earth may change significantly in some cases \citep[\textit{e.g.},][]{mostl2008,vourlidas2013,isavnin2014}. Nevertheless, the intrinsic flux rope type gives the first order approximation of a CME potential to drive a geospace disturbance and it is a critical input for several semi-empirical CME propagation models \citep[\textit{e.g.},][]{savani2015,savani2016,shiota2016,kay2016,isavnin2016}.

There is currently no practical method to measure the three-dimensional magnetic field in the corona in order to be able to determine the flux rope magnetic type in a CME. However, several morphological patterns in various solar phenomena associated with a CME eruption and its source region can be used as indirect proxies of the magnetic chirality of the resulting flux rope. These methods are based on soft X-ray and extreme-ultraviolet (EUV) observations of the pre-eruptive structure (which could be a flux rope or a sheared arcade), characteristics of the possible filament and/or flare association, and the evolution of the source active region. In addition, the observations of the polarity inversion line (PIL) over which the CME arises and/or the post-eruption arcades (PEAs) form can serve as proxies of the axial tilt of the CME flux rope and the axial field orientation. We will describe these techniques in more detail in the upcoming sections. 

So far, only a few studies have attempted to estimate the full CME flux rope type from remote-sensing observations \citep[\textit{e.g.},][]{marubashi1986,mcallister2001,yurchyshyn2001,mostl2008}. These previous studies have focused mostly on events that were associated with quiescent filaments or well defined active region filaments, and hence, could use filament characteristics to determine the chirality sign and the direction of the axial field \citep{marubashi1986,mcallister2001}. However, the majority of CMEs originate from active regions \citep{subramanian2001}, where filaments are typically smaller and less well-defined than quiescent filaments. Moreover, an active region does not always contain a filament. The study by \citet{subramanian2001}, near the rising phase of Solar Cycle 22, showed that 15\% of the CMEs studied originated from quiescent filaments, 44\% from active region filaments, and 41\% from active regions with no filament eruption.

In this article we perform two detailed case studies to determine the intrinsic magnetic structure of active region CMEs. We analyse data from multiple spacecraft and ground-based observatories to form a synthesis of several state-of-the-art remote sensing analysis techniques. In Section 2 we describe the data used and in Section 3 we discuss in more detail the different methods that can be applied to determine the flux rope chirality, axial tilt, and axial field orientation. The techniques we use have been previously validated in the literature, but they are currently in fragmented use and their combined potential to estimate the magnetic field of a CME is not yet fully explored. As we will demonstrate in this article, to systematically predict the flux rope type for active region CMEs one has to have several alternative proxies to determine the key flux rope properties. In Section 4 we apply our methods to two CME events and also make a first order validation of our results by comparing with \textit{in situ} observations of the corresponding CMEs. Finally, in Section 5 we discuss and summarize our results.

\section{Data Selection and Instruments}
\label{s:data}

We select our two case studies from the interplanetary CME list\footnotemark[1], compiled and maintained by Nieves-Chinchilla at NASA, and by examining solar and coronal (white-light) observations with the help of the \textit{Solar and Heliospheric Observatory} (SOHO) \textit{Large Angle Spectroscopic Coronagraph} (LASCO) CME Catalog\footnotemark[2], generated and maintained at the Coordinated Data Analysis Workshops (CDAW) data center by NASA and the Catholic University of America in cooperation with the Naval Research Laboratory. The selected events show flux rope signatures \textit{in situ} and they have a unique CME association, \textit{i.e.}, there were no multiple wide CMEs within a suitable time window that could have arrived to L1.

\footnotetext[1]{\url{http://wind.nasa.gov/index_WI_ICME_list.htm}}
\footnotetext[2]{\url{http://cdaw.gsfc.nasa.gov/CME_list/}}

To find the proper association between the interplanetary CME and the eruption from the solar atmosphere, we track the \textit{in situ} flux rope backwards in time to the Sun assuming a constant speed (given in the Nieves-Chinchilla list) and radial propagation. We search for associations within a two-day time window centered on the estimated CME onset time. As we look for Earth-directed events, we consider CMEs that have a wide angular span as listed in the
LASCO catalog (angular width $\gtrsim 120^{\circ}$).
Data from the \textit{Solar Terrestrial Relations Observatory} (STEREO) spacecraft \citep{kaiser2008} are used to confirm which CMEs are indeed Earth-directed. Our case studies occur in 2011 and 2012, when the two STEREO spacecraft were between 90$^{\circ}$ and 120$^{\circ}$ from the Sun-Earth line. This means that Earth-directed CMEs will be seen leaving the east limb in STEREO-A and the west limb in STEREO-B. To confirm that we connected the correct pair, we check the CME travel time using the empirical CME propagation model by \citet{gopalswamy2000}, using the linear (plane-of-sky) speed reported in the LASCO catalog. To associate the selected CMEs with the correct source region from which they erupted, we use well-known CME signatures that are observed in EUV, soft X-ray, and H$\alpha$ data, \textit{i.e.}, flares, post-eruptive arcades, flare ribbons, coronal EUV dimmings (transient coronal holes), and dark and cool rising material (signature of filament eruptions). 

EUV images and line-of-sight magnetograms taken with the \textit{Atmospheric Imaging Assembly} \citep[AIA:][]{lemen2012} and the \textit{Helioseismic and Magnetic Imager} \citep[HMI:][]{scherrer2012} onboard the \textit{Solar Dynamics Observatory} \citep[SDO:][]{pesnell2012} are used. SDO was launched on 11 February 2010 and has been operating since then in an inclined circular geosynchronous orbit. AIA takes images that span at least 1.3 solar diameters in multiple wavelengths nearly simultaneously, at a spatial resolution of 0.6 arcsec and at a cadence of 12 seconds. HMI creates full-disk magnetograms using the 6173 \AA \, spectral line with a spatial resolution of 0.5 arcsec and a temporal resolution of 45 seconds.

Soft X-ray data are supplied by the \textit{X-Ray Telescope} \citep[XRT:][]{golub2007} onboard \textit{Hinode} \citep[Solar-B:][]{kosugi2007}. \textit{Hinode} was launched on 22 September 2006 and has been operating since then in a nearly circular Sun-synchronous polar orbit around the Earth. XRT has various focal plane analysis filters, analyzing X-ray emission in a wide temperature range (from 1 to 10 MK). It provides two-arcsecond resolution images.

H$\alpha$ (6563 \AA) observations are from the \textit{Global Oscillations Network Group} (GONG). GONG is a six-station network of ground-based observatories located around the Earth to obtain nearly continuous observations of the Sun. The six observing sites are: the Big Bear Solar Observatory in California, USA, the High Altitude Observatory at Mauna Loa in Hawaii, USA, the Learmonth Solar Observatory in Western Australia, the Udaipur Solar Observatory in India, the Observatorio del Teide in the Canary Islands, and the Cerro Tololo Interamerican Observatory in Chile.

\textit{In situ} measurements are taken from the \textit{Wind} satellite, launched in November 1994 and operating close to L1 since 2004. We use the data from the \textit{Wind Magnetic Fields Investigation} \citep[MFI:][]{lepping1995} and the \textit{Wind Solar Wind Experiment} \citep[SWE:][]{ogilvie1995}, which provide 60-second and about 90-second resolution data, respectively.

\section{Research Methods}
\label{s:methods}

\subsection{Magnetic Structure of the Erupting Flux Rope }
\label{subs:magneticprior}

As discussed in the Introduction, there is currently no practical method to measure the magnetic field in the corona, which is needed in order to directly determine the flux rope type when it leaves the Sun. However, several proxies exist that can be used to achieve this goal. Here, we summarize the indirect proxies as presented in the literature. These proxies allow the identification of the flux rope chirality, tilt and axial field direction. In the following sections we will combine these proxies to estimate the flux rope type for our two case studies. Note that most of the methods described below are independent on whether the pre-eruptive structure is a sheared arcade or a flux rope.

\subsubsection{Chirality of the Flux Rope}
\label{subs:chirality}

To estimate the flux rope chirality we carefully analyze the source active region and the evolution of the erupting structure. It is expected that the CME flux rope has the same chirality as the source region in which it formed, since magnetic helicity is a conserved quantity even during magnetic reconnection \citep{berger2005}. The methods to estimate the chirality are: 

1) \textbf{Magnetic tongues.} The global chirality of an active region can be estimated by analyzing the line-of-sight magnetograms during the active region's emergence phase. Active regions form from emerging twisted flux tubes ($\Omega$-loops). When the apex of such tubes crosses the photosphere, the vertical projection of the azimuthal component of the field manifests itself in the magnetogram data as ``magnetic tongues'' \citep{lopezfuentes2000,luoni2011}. Magnetic tongues are elongations of the main polarities, where a positive twist is shown by the leading magnetic polarity extending under the southern edge of the trailing polarity and a negative twist is represented by its mirror image (magnetic tongues are a polarity-invariant chirality proxy). 

2) \textbf{Filament details.} If the CME is associated with a filament eruption, the chirality of the flux rope can be deduced from studying the detailed structure of the filament before the CME onset. Sinistral (dextral) filaments are embedded in regions of positive (negative) chirality \citep{martin1996,martin2003}. The sinistral or dextral nature of filaments can be revealed by various patterns in H$\alpha$ observations, \textit{e.g.}, by the bearing of the filament legs, the orientation of the fibrils in filament channels and the orientation of filament barbs with respect to the filament axis \citep[\textit{e.g.},][]{martin1994,martin1998}. Moreover, for positive (negative) chirality the filament apex rotates clockwise (counterclockwise) upon eruption \citep{green2007,lynch2009}. Filaments can also be studied at EUV wavelengths and the chirality can be argued from the geometry of the crossings between emission and absorption threads \citep{chae2000}. 

3) \textbf{X-ray and/or EUV Sigmoids.} Sigmoids are S-shaped soft X-ray or EUV emission structures that can be considered as coronal tracers of a flux rope \citep{rust1996, canfield1999,green2009,green2014}. The S-shaped emission structure is formed by field lines threading quasi-separatrix layers associated with a flux rope embedded in an arcade \citep{titov1999}. A sigmoid can have one of two orientations depending on the chirality of the magnetic field in the region where it forms \citep[\textit{e.g.,}][]{pevtsov1997,green2007}: forward (reverse) S-sigmoids form in the regions dominated by positive (negative) chirality. 

4) \textbf{Skew of the coronal loops.} An additional soft X-ray and/or EUV feature that can be used as a proxy of the chirality is the skew, \textit{i.e.} the acute angle that the coronal loops overlying the pre-eruptive flux rope or sheared arcade make with the PIL or the filament axis. The loops are defined as left-skewed or right-skewed according to the sense of the arcade loops crossings over the filament or filament channel \citep{mcallister1998,martin2012}. Left- (right-) skewed arcades are associated to dextral (sinistral) filaments and negative (positive) helicity flux ropes \citep{martin1998}.

5) \textbf{Flare ribbons.} The observational signature of the energy release within quasi-separatrix layers during a solar flare is the brightening of two J-shaped flare ribbons. Two reverse J-shapes indicate negative chirality and forward J-shapes indicate positive chirality. Also the orientation and displacement of the ribbons along the PIL reflect the sign of twist in the flux rope (this is an indication of the remaining magnetic shear present after reconnection). If the PIL is vertical on the solar image, the left ribbon is displaced downwards and the right ribbon upwards for positive chirality, while the situation reverses for negative chirality \citep{demoulin1996}. 

6) \textbf{Hemispheric helicity rule}: There is a tendency for magnetic structures on the Sun to have negative (positive) helicity in the northern (southern) hemisphere.  Such a pattern is known as the ``hemispheric helicity rule'' \citep{pevtsov2003}. \citet{bothmer1998} used this general hemispheric rule to relate the flux rope properties \textit{in situ} to the properties of their source region. While such rules can be powerful in a statistical sense, their use as a reliable proxy of the magnetic characteristics of individual CMEs is limited since the hemispheric helicity rule only holds true in around 60-75\% of emerging active regions \citep[][]{pevtsov2014}.

\subsubsection{Flux Rope Tilt and Axial Field Orientation}
\label{subs:magneticeruption}

The orientation of an erupting flux rope (\textit{i.e.}, the orientation of its axis) can be considered to be more or less parallel to the orientation of the PIL in the solar source region \citep{marubashi2015} or to the orientation of the post-eruption arcades (PEAs) \citep{yurchyshyn2008}. PEAs are often the clearest signature of the CME eruption in the low corona and they are visible in both soft X-ray \citep{mcallister1996,hudson1997} and EUV \citep{tripathi2004} observations. These arcades are associated with reconnection that occurs in the wake of an erupting CME. 

We determine the PIL location and orientation by eye, \textit{i.e.} we determine the location where the polarity of the magnetic field reverses and approximate it with a straight line. We define $|\tau|$ as the absolute value of the angle within the range $\pm 90^{\circ}$ that the PIL makes with the solar ecliptic, assuming that the flux rope at its nose is perpendicular to the Sun-Earth line ($-\mathbf{\hat{x}}_{GSE}$, in Geocentric solar ecliptic (GSE) coordinates). We approximate that $|\tau| < 45^{\circ}$ corresponds to bipolar (parallel) flux ropes \textit{in situ}, while $|\tau| > 45^{\circ}$ corresponds to unipolar (perpendicular) flux ropes \textit{in situ}. 

The direction of the flux rope axial field can be taken to be the direction of the magnetic field that runs nearly parallel to the PIL, which depends on the helicity sign of the source region \citep{wang2013,marubashi2015}. This direction can be argued from photospheric magnetogram data and the coronal configuration: the field is directed left (right) when looking from the positive magnetic polarity side along the PIL for a positive (negative) helicity source region \citep{bothmer1994,marubashi2015}. 

As a further confirmation of the axial field direction, we study base-difference images of the source region during the rise time of the flux rope \citep[\textit{e.g.},][]{mandrini2005}. The key here is to locate reliably the footpoints of the flux rope. One viable method is provided by EUV dimmings \citep{hudson1997} that correspond to the evacuation of coronal material that is fed into the rising CME \citep{hudson1997} and that are generally believed to map the footpoints of the CME in the corona \citep{thompson2000}. Hence, we search for signs of EUV dimmings in base-difference images and overlay the dimming regions onto line-of-sight magnetogram data to determine in which magnetic polarities the flux rope is rooted. Then, the axial field is directed from the positive footpoint to the negative one.

After obtaining the chirality of the source region, the tilt of the flux rope, and its axial field direction, we make our prediction of the flux rope type.

\subsection{Magnetic Structure \textit{in situ}}
\label{subs:magneticinsitu}

We apply the minimum variance analysis technique (MVA, \citealp{sonnerup1967}) to the \textit{in situ} data to estimate the orientation of the flux rope axis at 1 AU (latitude $\theta_{A}$ and longitude $\phi_{A}$ in angular coordinates) and to verify the coherent rotation of the magnetic field vectors. The flux rope axis corresponds to the intermediate variance direction, where $\theta_{A}=90^{\circ}$ is defined northward and  $\phi_{A}=90^{\circ}$  is defined eastward. The latitude can then be used to estimate the inclination of the axis with respect to the ecliptic. The consistency of the MVA method is taken into account by checking that $\lambda_{2}/\lambda_{3} \geq 2$ \citep[\textit{e.g.},][]{lepping1980,bothmer1998,huttunen2005}, where $\lambda_{2}$ and $\lambda_{3}$ are the intermediate and minimum eigenvalues, respectively.
In addition, we estimate the crossing distance of the spacecraft from the apex of the CME flux rope loop with the angle, $\alpha$, that the shock normal makes with the radial direction, \textit{i.e.} the Sun-Earth direction, $-\mathbf{\hat{x}}_{GSE}$ \citep[\textit{e.g.},][]{janvier2015,palmerio2016}. $\alpha \approx 0$ means that the spacecraft crosses the CME close to its apex and the angle increases as the crossing takes place more on the flank of the CME. The shock normals are obtained from the Heliospheric Shock Database\footnotemark[3], developed and maintained at the University of Helsinki.

\footnotetext[3]{\url{http://ipshocks.fi/}}

As a proxy for the spacecraft crossing distance from the center of the flux rope, we estimate the impact parameter using the total perpendicular pressure, $P_{\perp}$, defined as the sum of the magnetic pressure and the thermal pressure perpendicular to the magnetic field \citep[\textit{e.g.},][]{jian2006}. An interplanetary CME can present three different $P_{\perp}$ profiles, which results in three different groups. In group 1 $P_{\perp}$ has a central maximum in the magnetic obstacle, in group 2 $P_{\perp}$ has a plateau-like profile, and in group 3 $P_{\perp}$ increases rapidly and then gradually decreases \citep{jian2005}. For group 1 events the spacecraft crosses the flux rope centrally, while for group 2 and 3 the spacecraft crossing takes place at a larger distance from the axis. The perpendicular pressure is obtained from the Solar Wind Data service\footnotemark[4], maintained at the Space Science Center, University of California, Los Angeles.

\footnotetext[4]{\url{http://www-ssc.igpp.ucla.edu/forms/polar/corr_data.html}}

In addition, we apply the Grad-Shafranov reconstruction (GSR, \citealp{hau1999,hu2002}). The GSR gives estimates of the orientation (latitude $\Theta$ and longitude $\Phi$ in angular coordinates), chirality, impact parameter, and cross section of the flux rope. The advantage of this method is that it relaxes the force-free assumption and reconstructs the flux rope without a preset geometry, assuming only that the magnetic field has translational symmetry with respect to an invariant axis direction. We use a modification of this method described in \citet{isavnin2011}, and determine the flux rope invariant axis and the closest approach of the spacecraft to it by trial and error. For each possible axis direction, we project the magnetic field data onto the plane perpendicular to the axis, and calculate transverse pressure, $P_{t}$, and magnetic potential, $A$. The $P_{t}(A)$ curve forms two branches, corresponding to the trajectory of the spacecraft, first towards the flux rope axis and then away from it. The point in the curve that connects the two branches represents the closest approach of the spacecraft to the invariant axis. For each trial the residue between the two branches is calculated, and the results are displayed as a residual map. The direction with minimum residue corresponds to the estimated invariant axis direction.

\section{Results}
\label{s:results}  

\subsection{Event 1: CME on 2 June 2011}
\label{subs:june2011}

Our first case study describes a CME that erupted on 2 June 2011 between the two NOAA active regions (ARs) 11226 and 11227. The CME was detected \textit{in situ} two days later. The same CME association has been made by \citet{colaninno2013}. We first describe the coronal signatures of the event and then give a prediction for the magnetic flux rope type based on the remote-sensing observations, followed by an analysis of its \textit{in situ} signatures. 

\subsubsection{Coronal Observations}

The CME (halo) was first observed by LASCO C2 on 2 June at 08:12 UT, having a plane-of-sky linear speed of 976 km s$^{-1}$. The same event appeared in the STEREO A COR1 (emanating eastwards) and STEREO B COR1 (emanating westwards) field-of-view on 2 June at 07:45 UT. In addition, a very faint eruption was detected about an hour before the bright halo CME, appearing in the LASCO C2 field of view at 07:24 UT, and the STEREO A and B COR1 at 06:45 UT. This eruption (from now on Eruption 1) was very faint and slow (LASCO C2 reported a plane-of-sky velocity of 253 km s$^{-1}$), but nevertheless was related to the same eruptive event that formed the halo CME (from now on Eruption 2). Thus, even though we assume that only Eruption 2 reached the Earth, we discuss here both eruptions in order to get a clearer understanding of the event in its entirety.

\begin{figure}
   \vspace{0.05\textwidth}	

  \centerline{
               \includegraphics[width=0.27\textwidth,clip=]{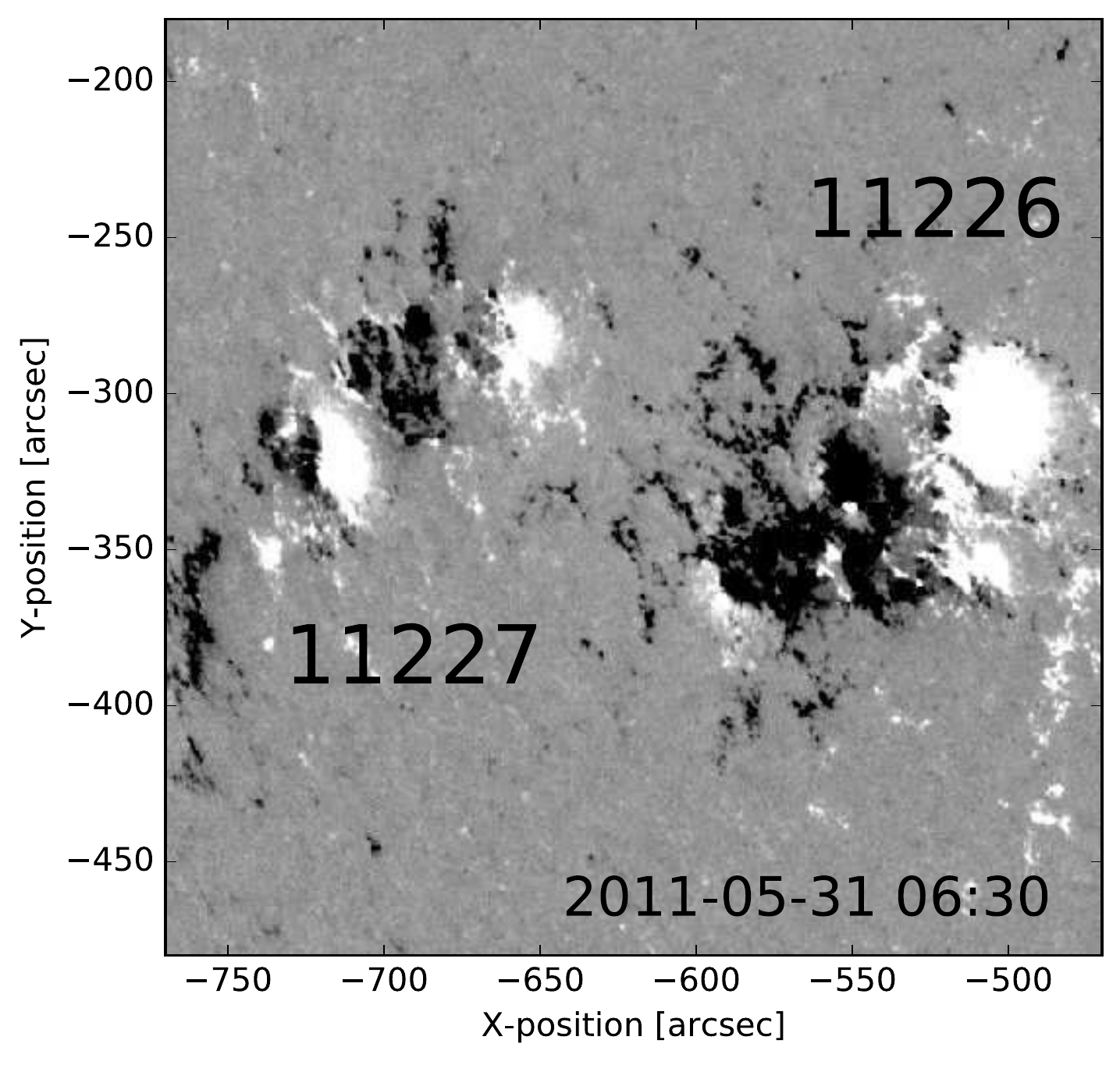}
               \includegraphics[width=0.235\textwidth,clip=]{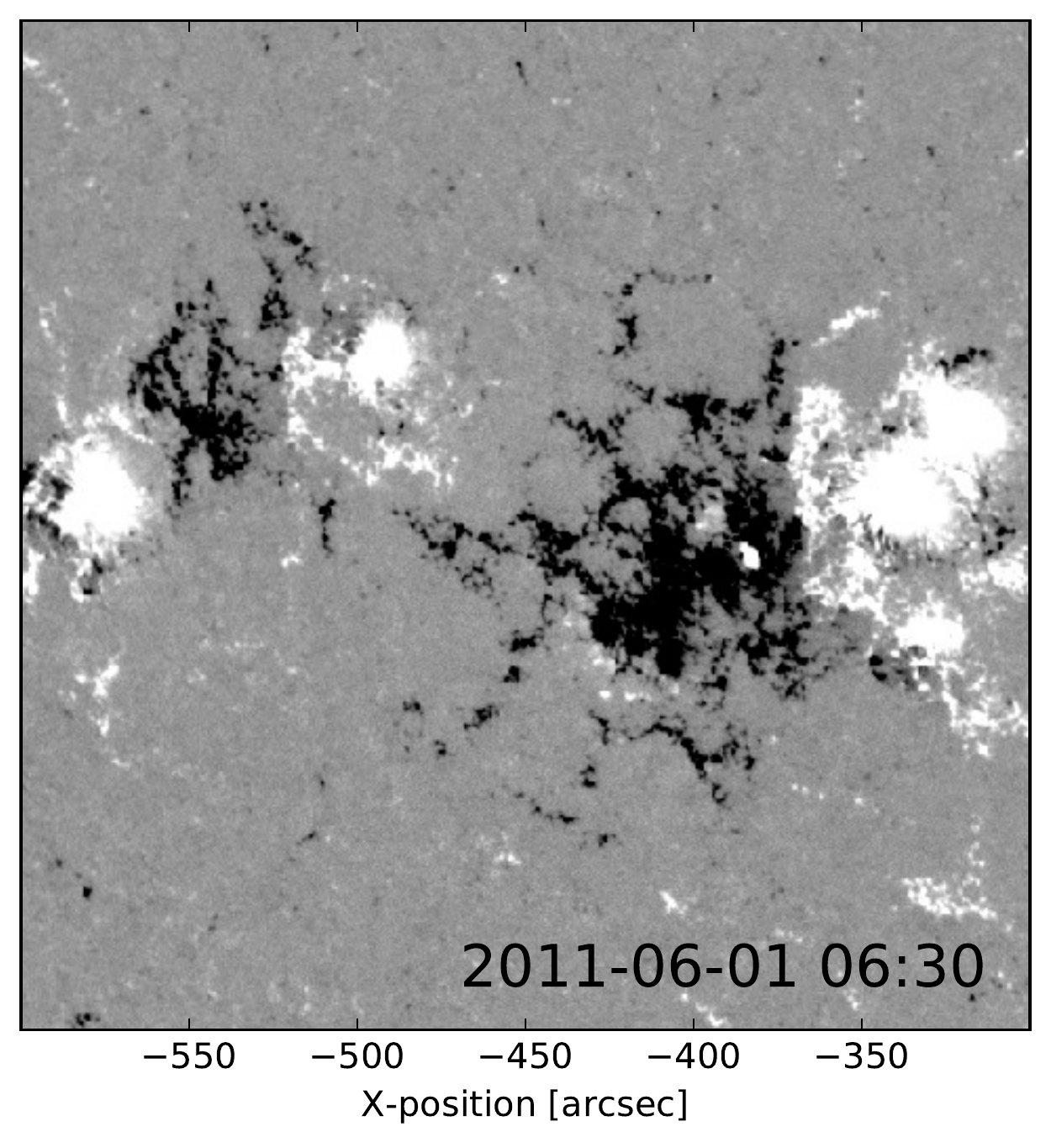}
			  \includegraphics[width=0.235\textwidth,clip=]{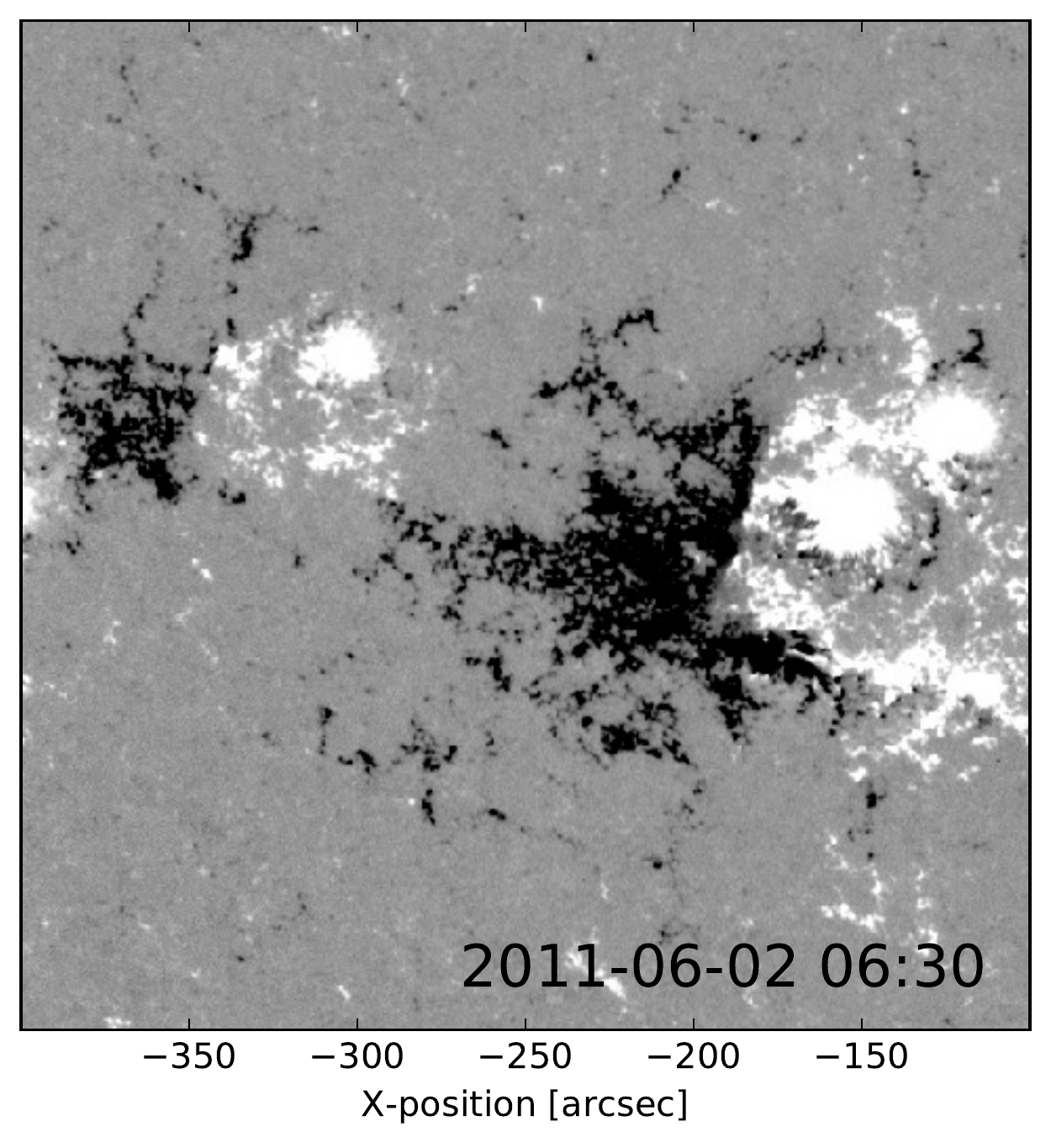}
               \includegraphics[width=0.235\textwidth,clip=]{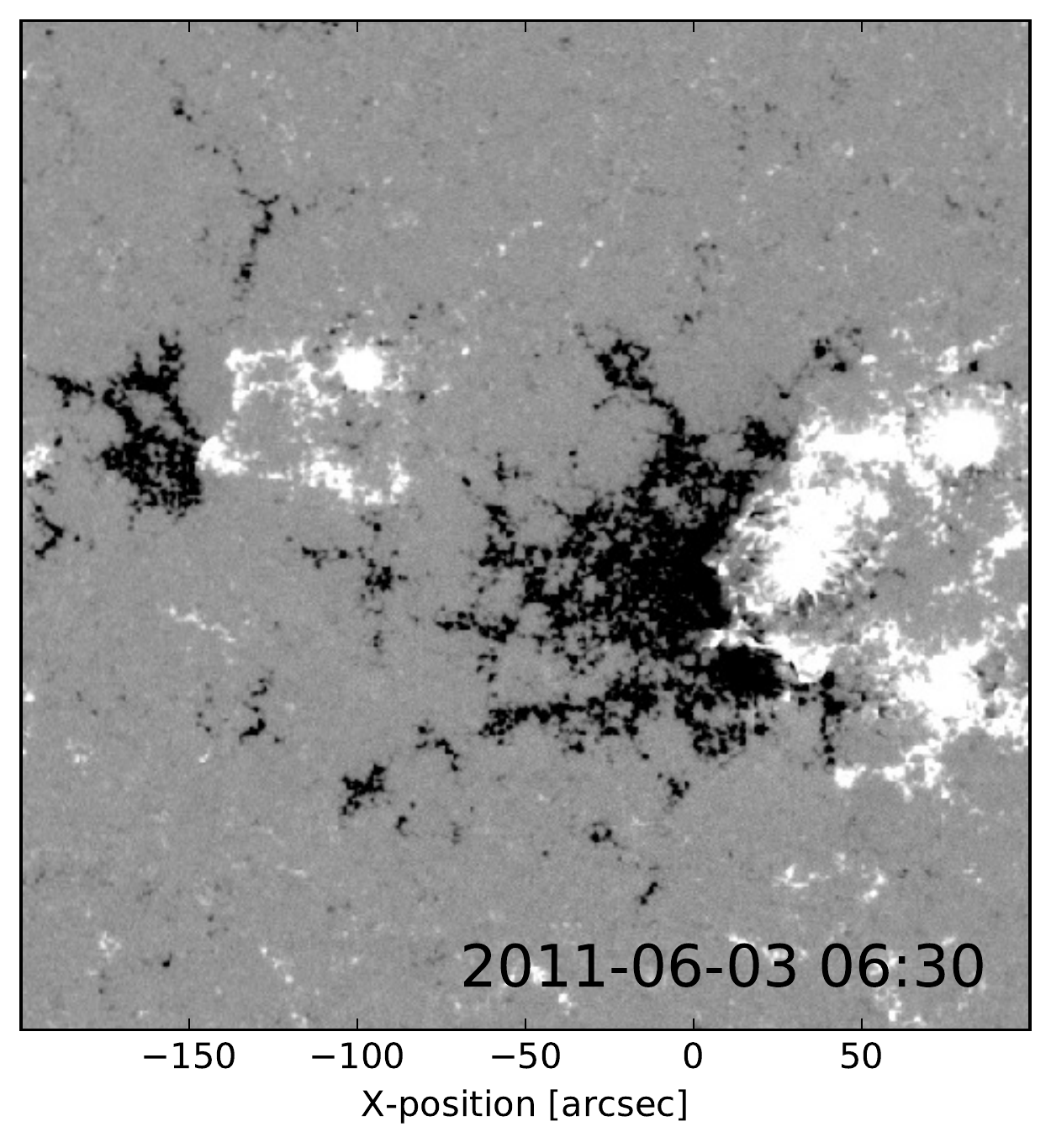}
			  }
			  
      \vspace{-0.31\textwidth}   
     \centerline{\large \bf                         
      \hspace{0.29 \textwidth} {HMI MAGNETOGRAM}
      	\hfill}
      	
      \vspace{0.017\textwidth}   
     \centerline{\normalsize \bf    
      \hspace{0.03 \textwidth}  \color{white}{(a)}
      \hspace{0.18 \textwidth}  \color{white}{(b)}
      \hspace{0.179 \textwidth}  \color{white}{(c)}
      \hspace{0.182\textwidth}  \color{white}{(d)}
         \hfill}
      	
     \vspace{0.30\textwidth}
 
 	\centerline{
               \includegraphics[width=0.27\textwidth,clip=]{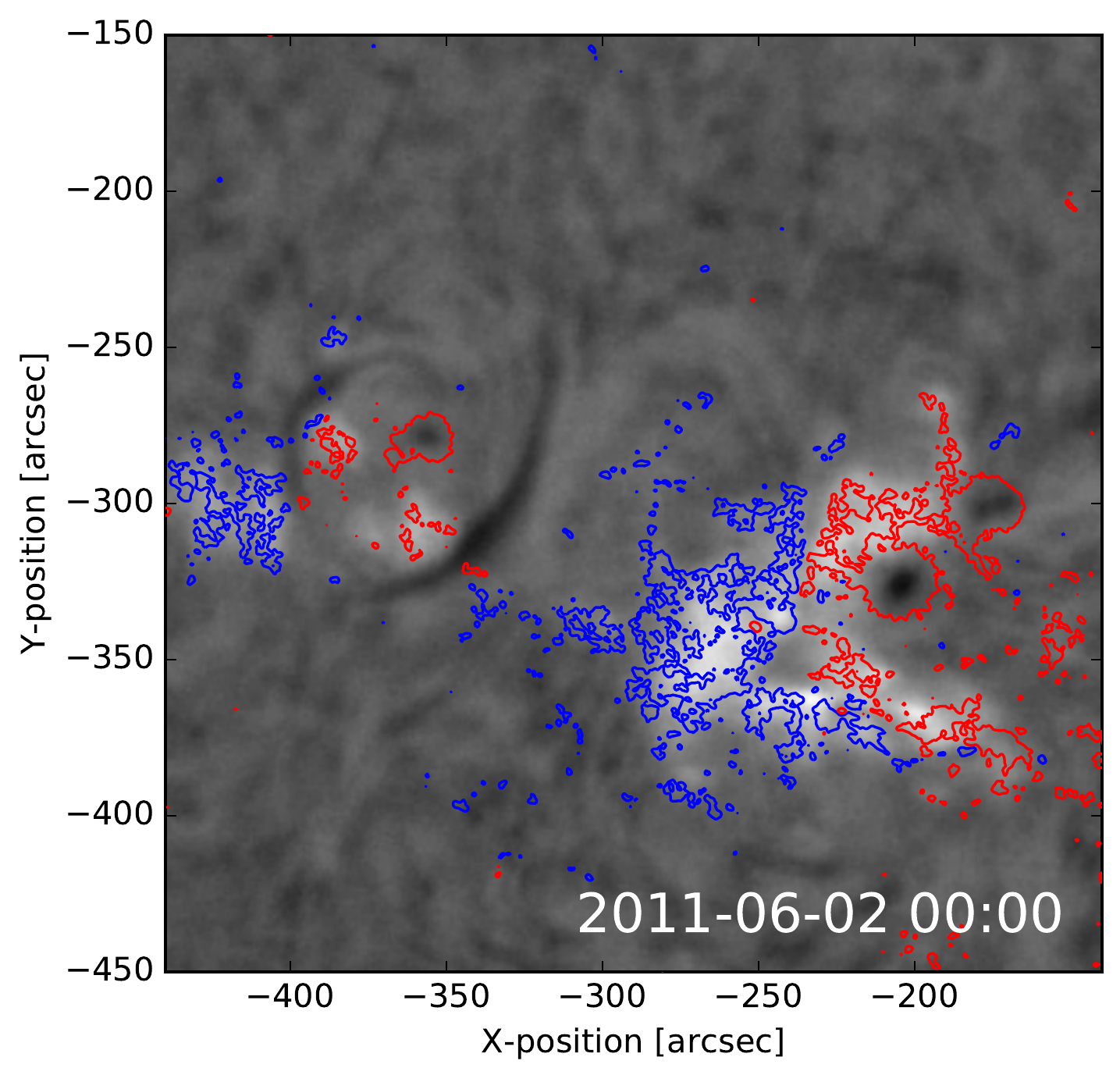}
               \includegraphics[width=0.235\textwidth,clip=]{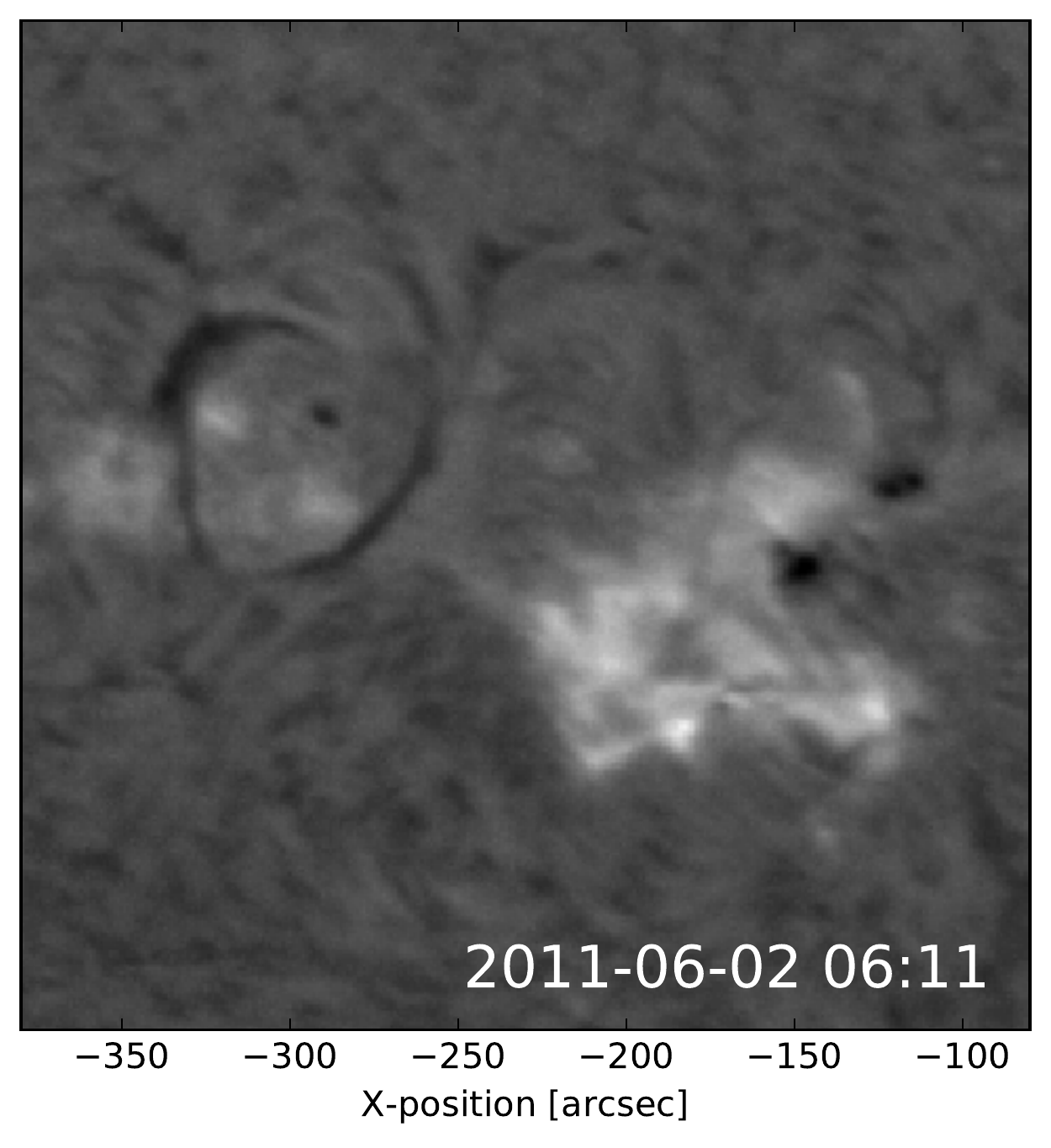}
			  \includegraphics[width=0.235\textwidth,clip=]{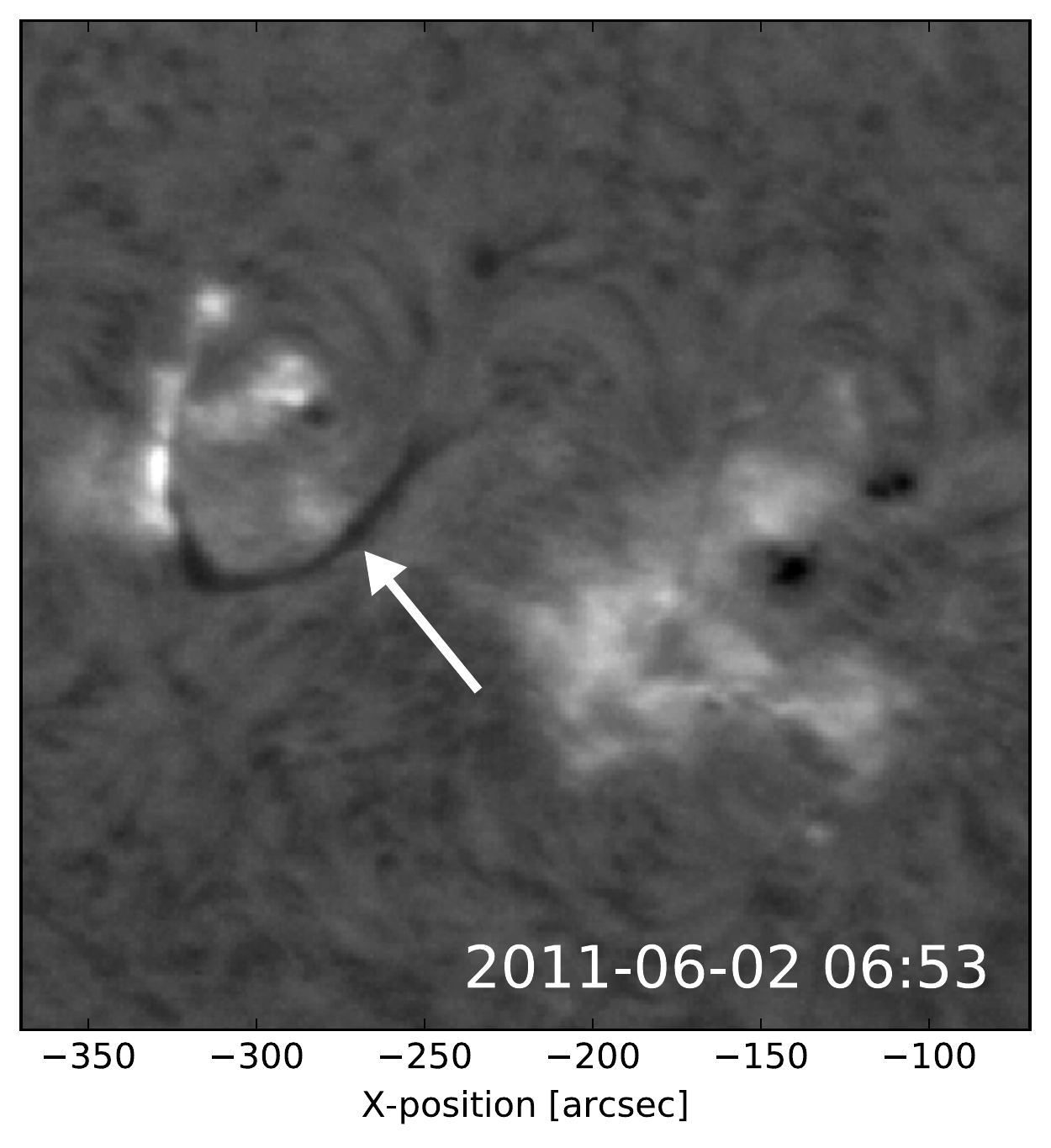}
               \includegraphics[width=0.235\textwidth,clip=]{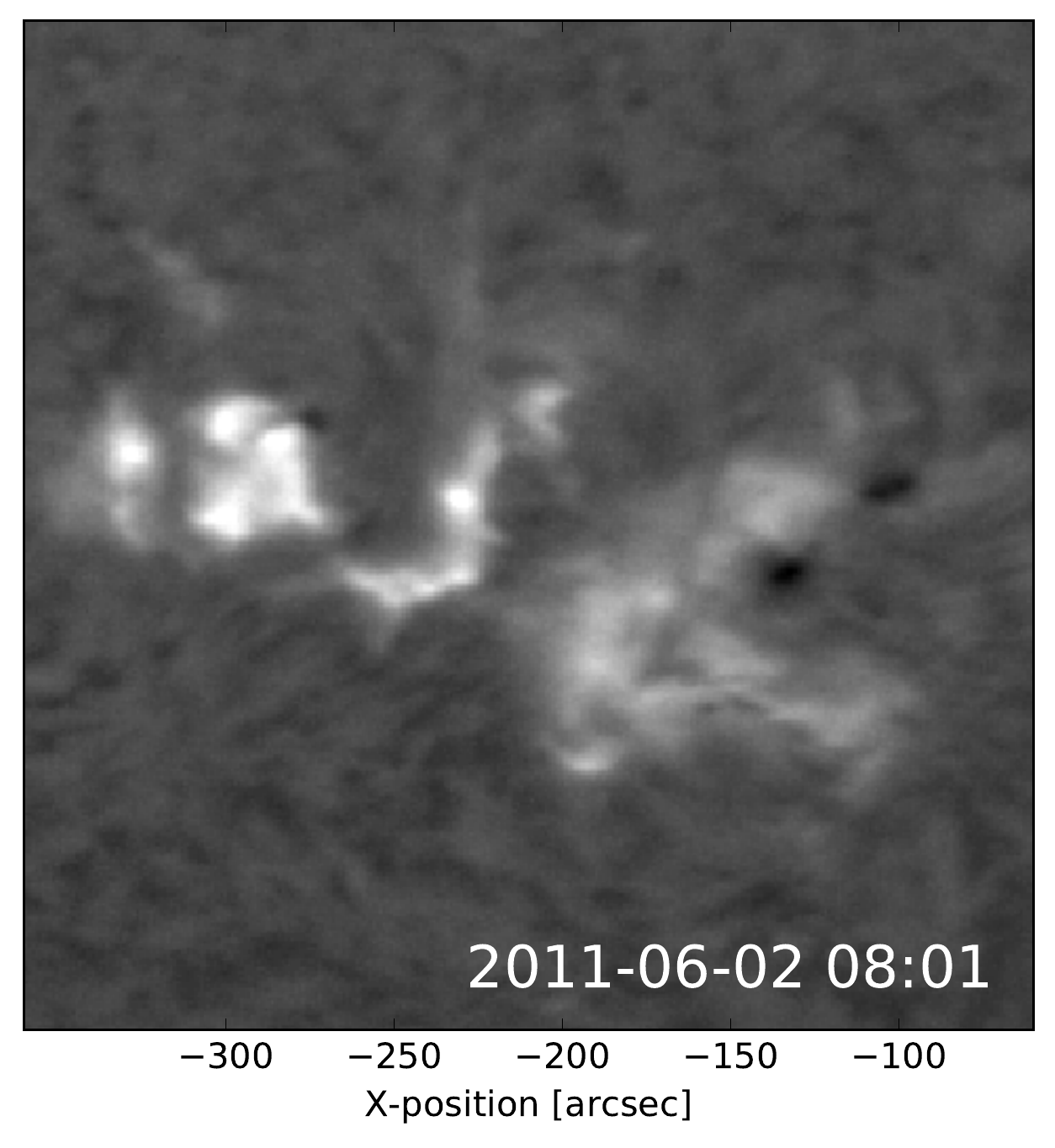}
			  }
              
      \vspace{-0.31\textwidth}   
     \centerline{\large \bf                         
      \hspace{0.42 \textwidth} {GONG H$\alpha$}
      	\hfill}
      	
      	\vspace{0.017\textwidth}   
     \centerline{\normalsize \bf    
      \hspace{0.03 \textwidth}  \color{white}{(e)}
      \hspace{0.182 \textwidth}  \color{white}{(f)}
      \hspace{0.183 \textwidth}  \color{white}{(g)}
      \hspace{0.181\textwidth}  \color{white}{(h)}
         \hfill}
      	
     \vspace{0.3\textwidth}
     
     \centerline{
               \includegraphics[width=0.27\textwidth,clip=]{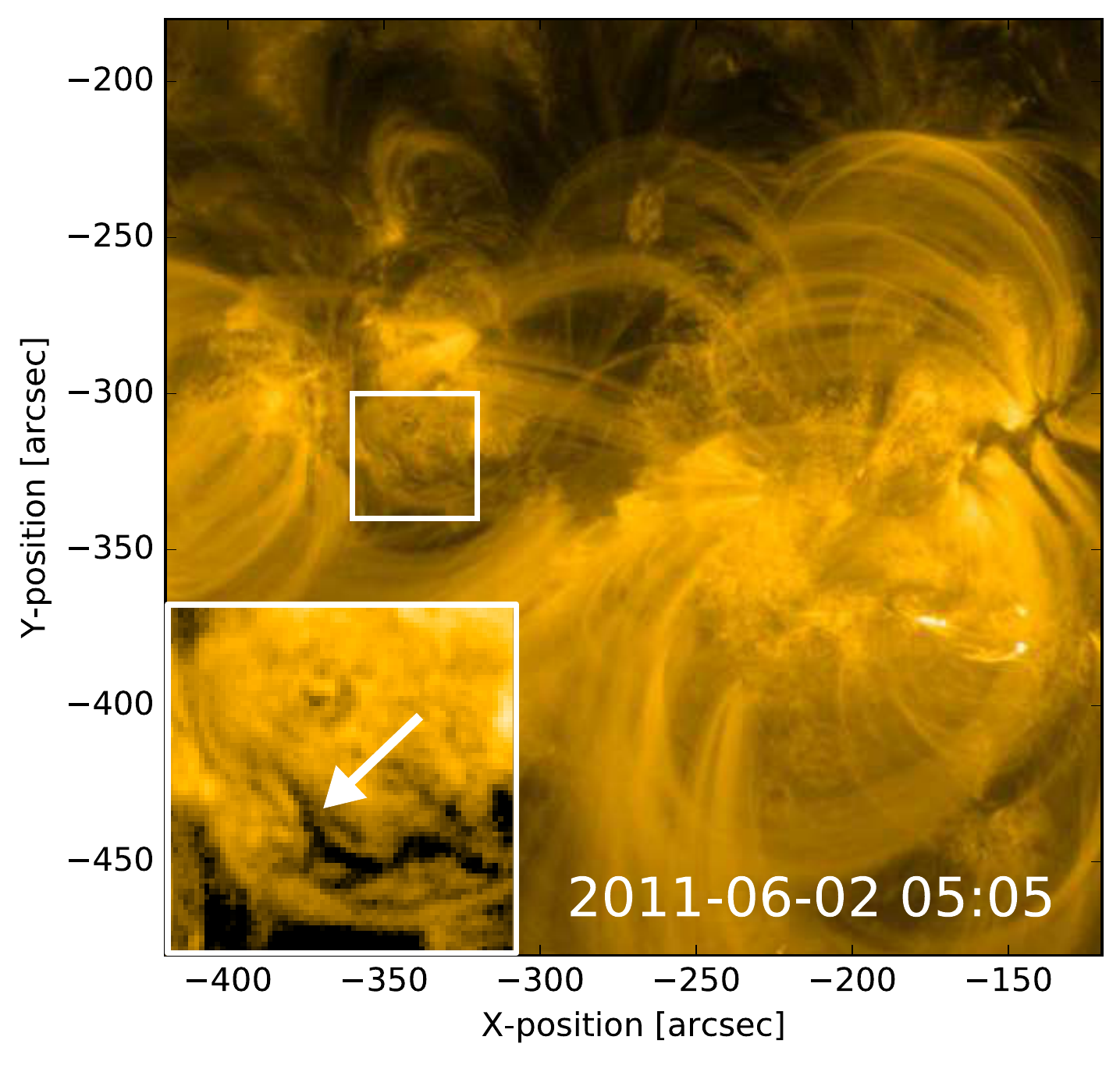}
               \includegraphics[width=0.235\textwidth,clip=]{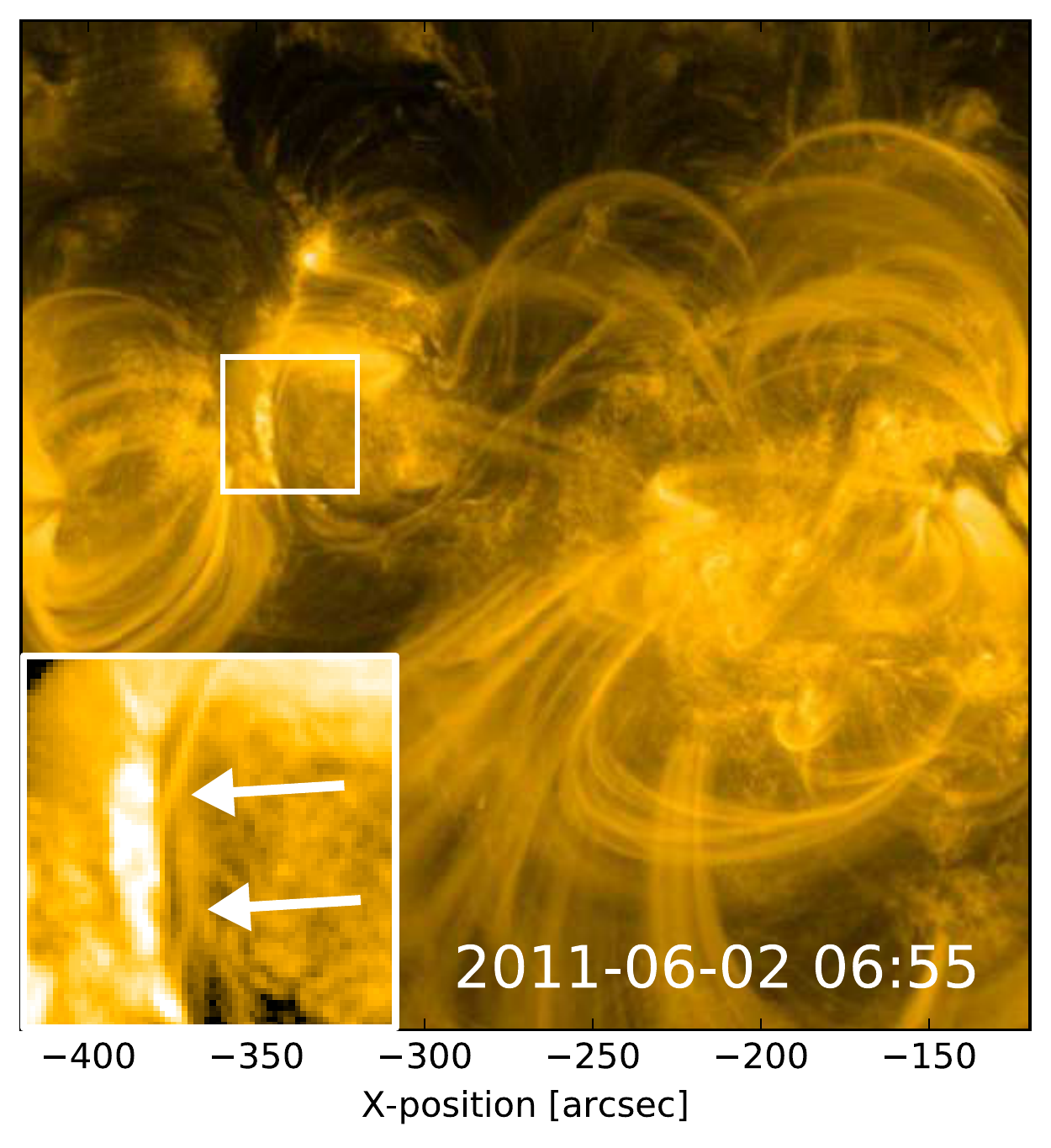}
			  \includegraphics[width=0.235\textwidth,clip=]{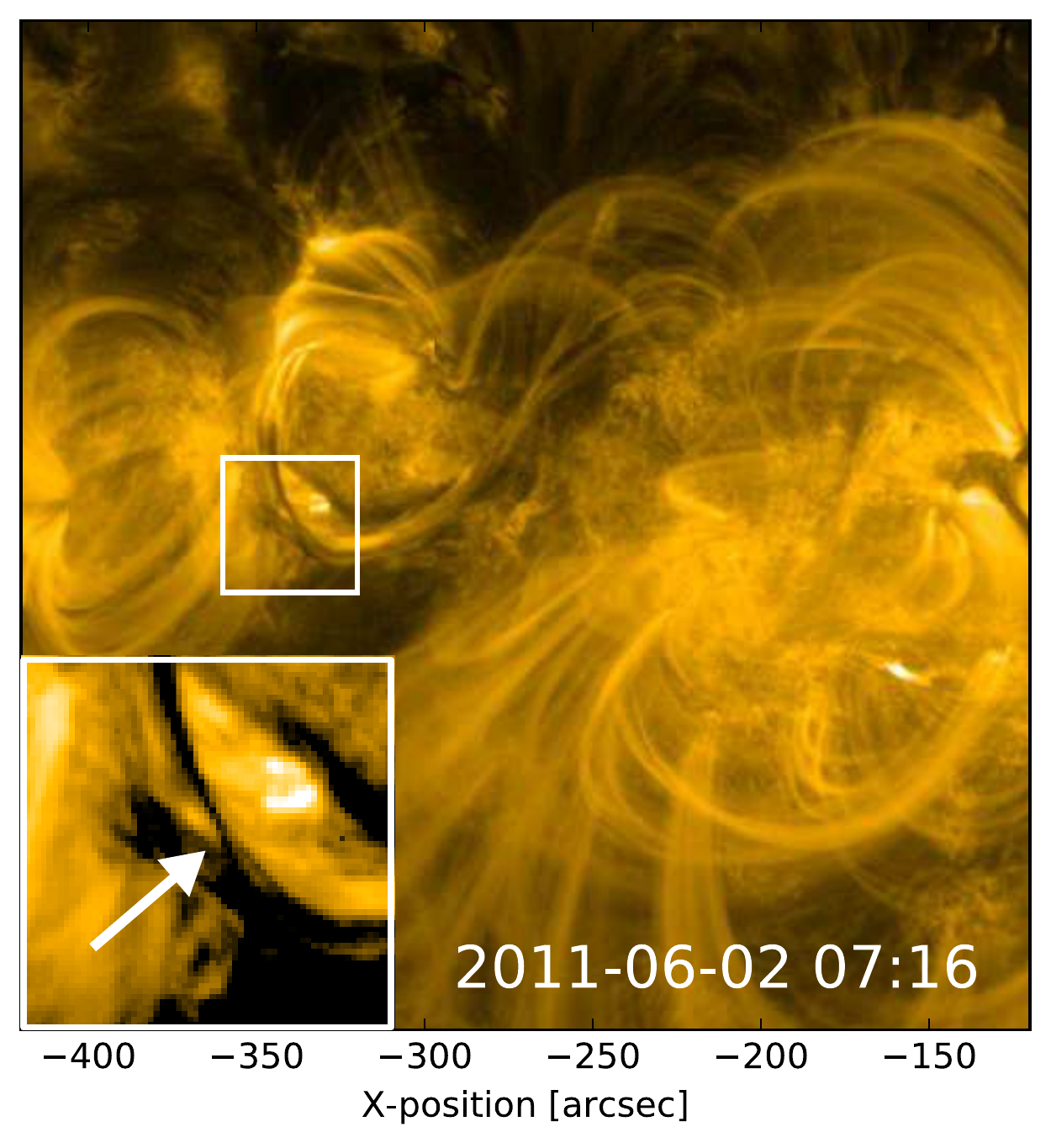}
               \includegraphics[width=0.235\textwidth,clip=]{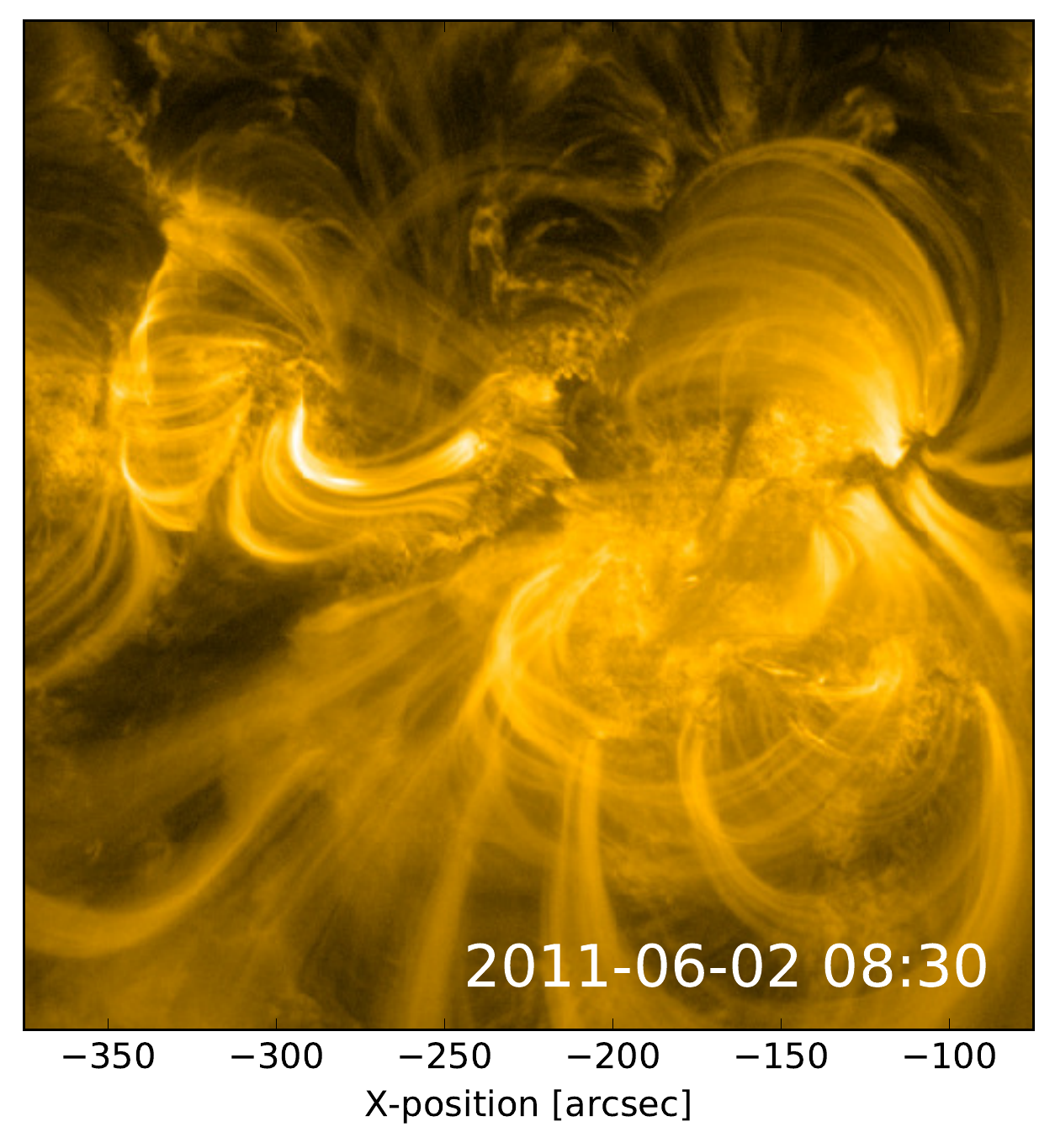}
			  }
              
      \vspace{-0.31\textwidth}   
     \centerline{\large \bf                         
      \hspace{0.36 \textwidth} {AIA EUV 171 \AA}
      	\hfill}
      	
      	\vspace{0.017\textwidth}   
     \centerline{\normalsize \bf    
      \hspace{0.03 \textwidth}  \color{white}{(i)}
      \hspace{0.187 \textwidth}  \color{white}{(j)}
      \hspace{0.187 \textwidth}  \color{white}{(k)}
      \hspace{0.18\textwidth}  \color{white}{(l)}
         \hfill}
      	
     \vspace{0.30\textwidth}
 
 	\centerline{
               \includegraphics[width=0.27\textwidth,clip=]{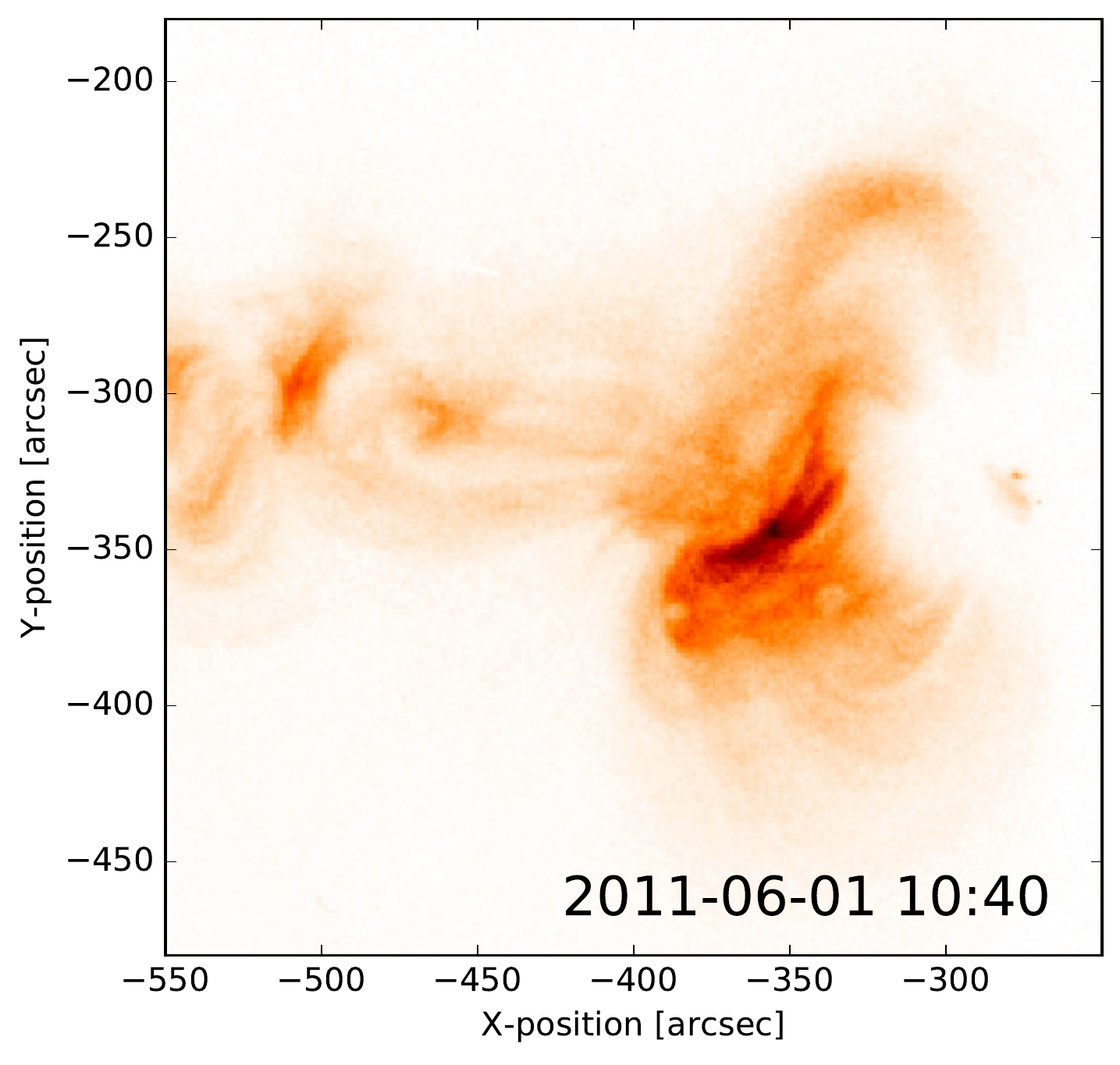}
               \includegraphics[width=0.235\textwidth,clip=]{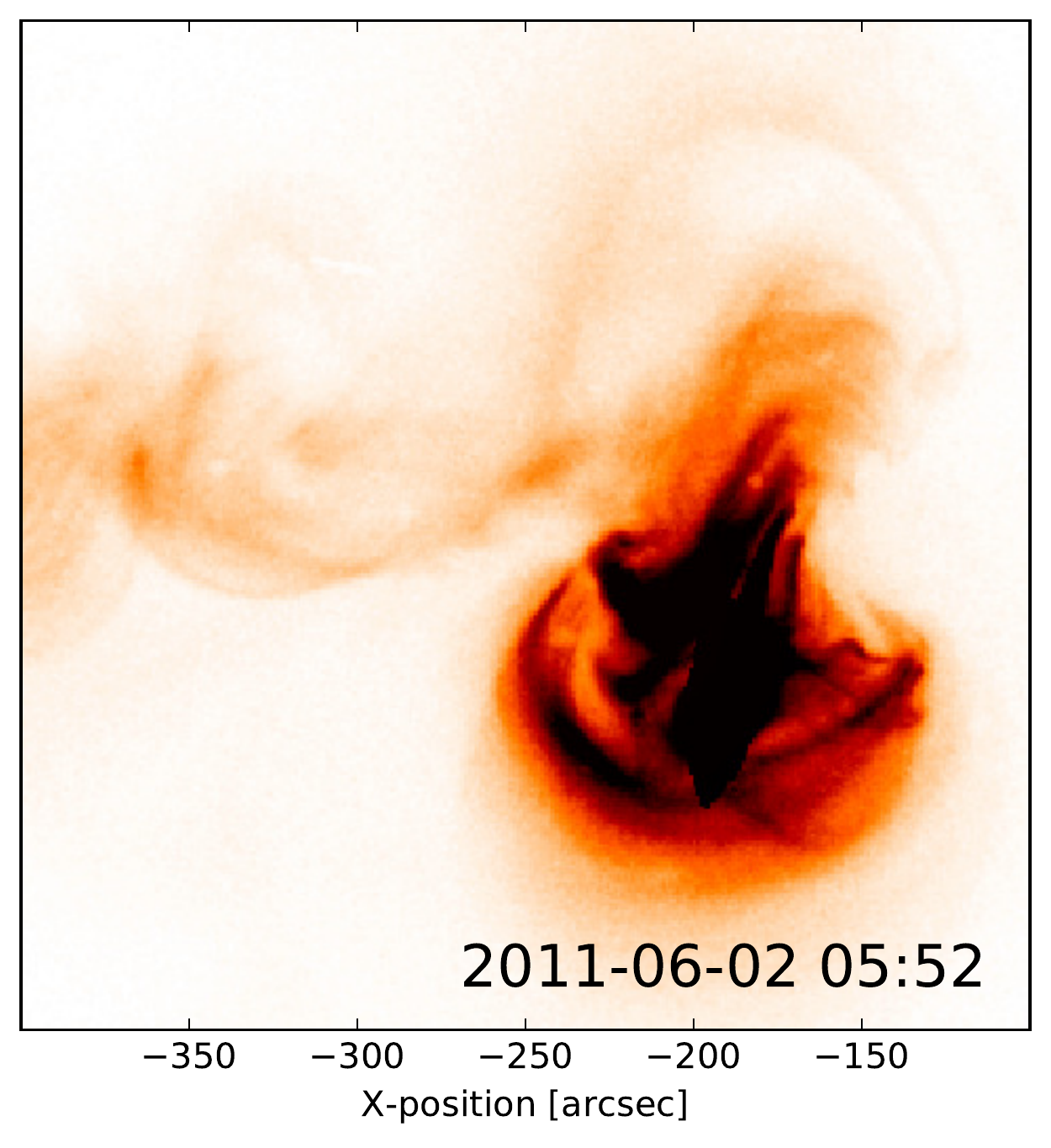}
			  \includegraphics[width=0.235\textwidth,clip=]{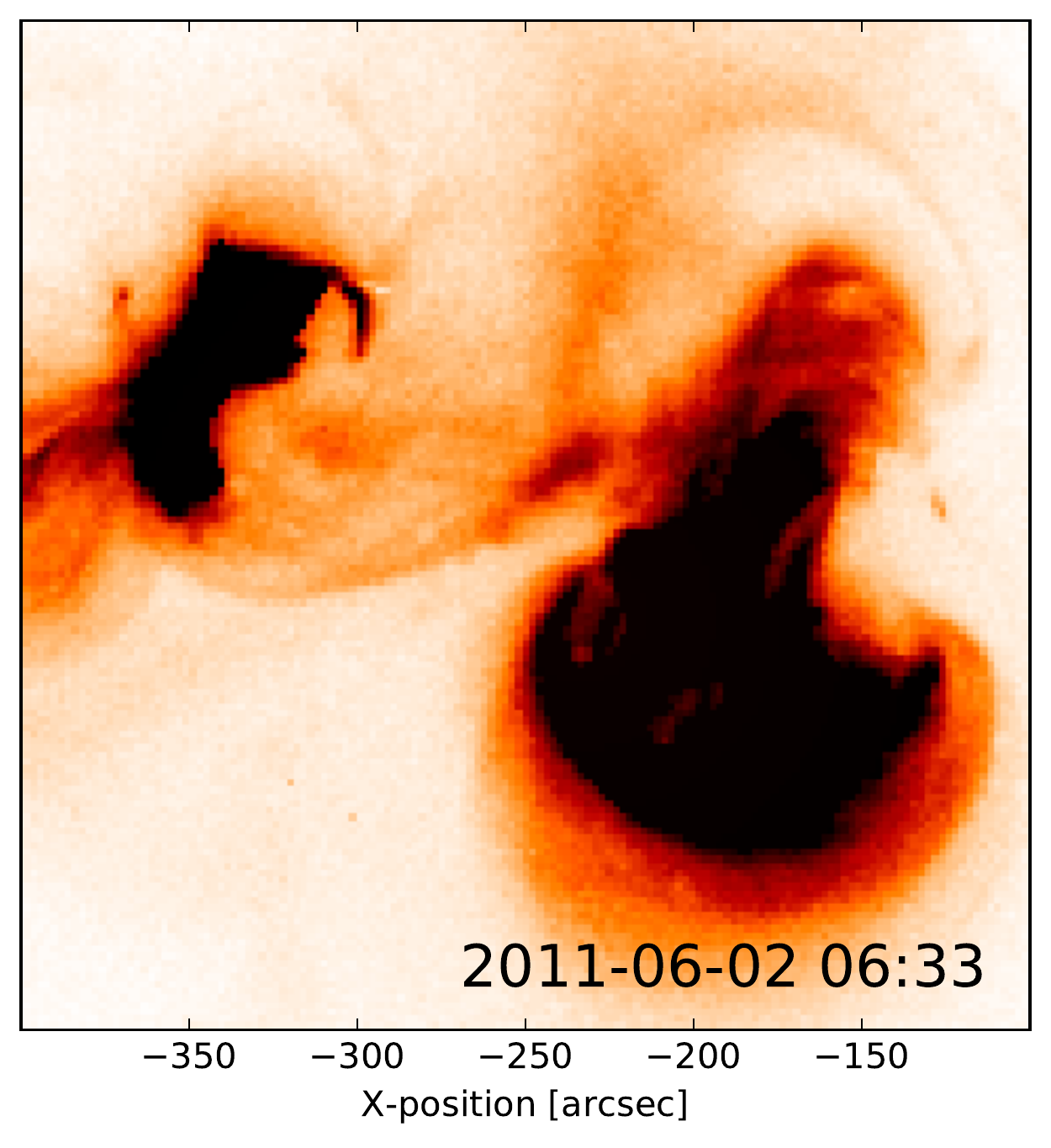}
               \includegraphics[width=0.235\textwidth,clip=]{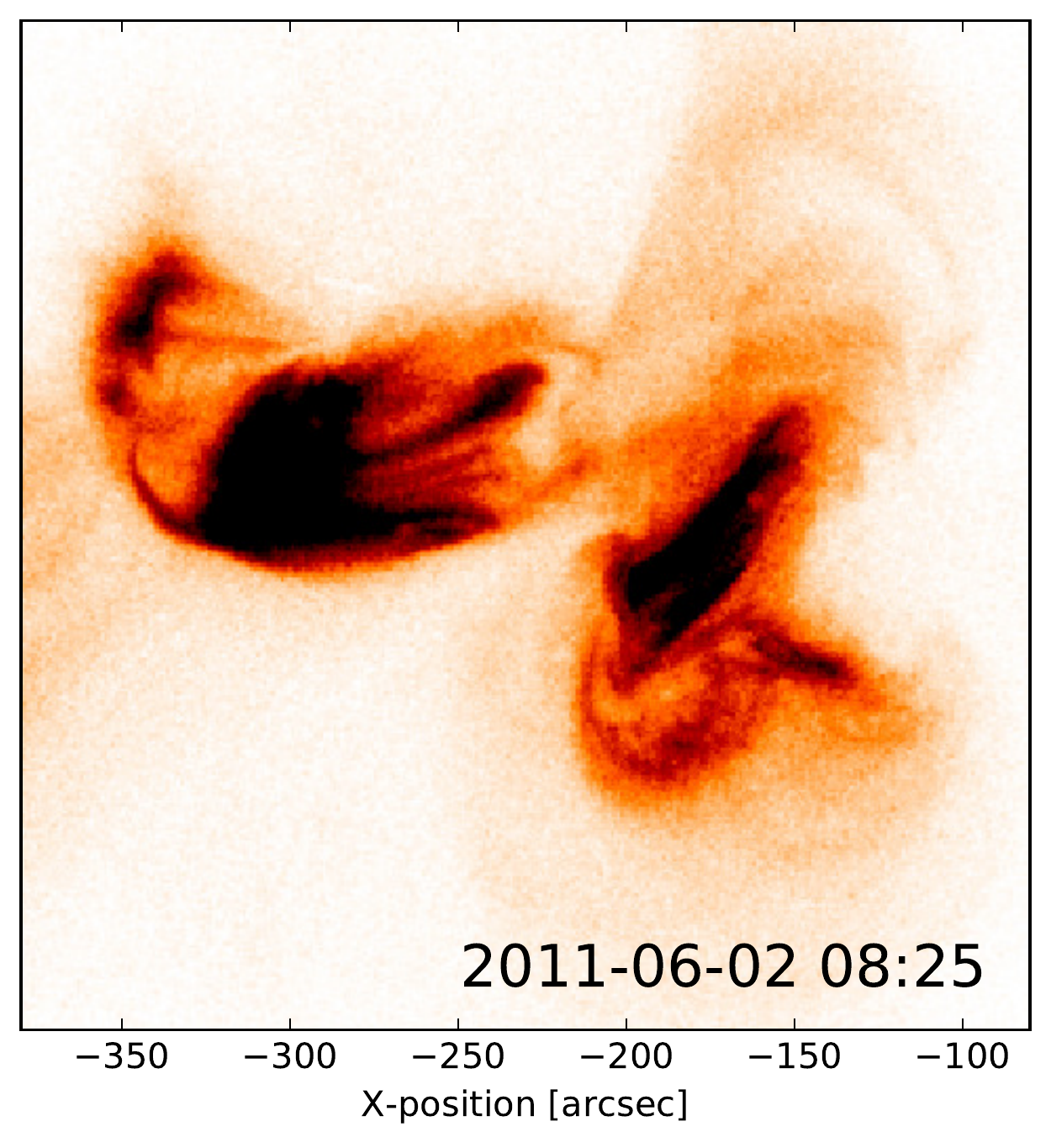}
			  }
              
      \vspace{-0.31\textwidth}   
     \centerline{\large \bf                         
      \hspace{0.33 \textwidth} {XRT SOFT X-RAYS}
      	\hfill}
      	
      	\vspace{0.017\textwidth}   
     \centerline{\normalsize \bf    
      \hspace{0.03 \textwidth}  \color{black}{(m)}
      \hspace{0.17 \textwidth}  \color{black}{(n)}
      \hspace{0.18 \textwidth}  \color{black}{(o)}
      \hspace{0.181\textwidth}  \color{black}{(p)}
         \hfill}
      	
     \vspace{0.25\textwidth}
     
              \caption{\textit{Top row:} Evolution of ARs 11226 and 11227 in line-of-sight magnetograms as observed by HMI/SDO and saturated to $\pm 200$ G. The third panel corresponds to the time of Eruption 1. 	
               \textit{Second row:} H$\alpha$ evolution of the eruptions as observed by the BBSO (first panel) and the Udaipur (remaining three panels) observatories from GONG. The first panel has been overlaid with SDO magnetogram contours saturated to $\pm 200$ G (blue is used for the negative polarity and red for the positive polarity). The arrow in the third panel indicates the part of the filament involved in Eruption 2.
			 \textit{Third row:} Evolution in EUV of the eruptive event during 2 June, as observed by AIA/SDO. The images are taken with the 171 \AA \, filter. Crossings of dark and bright filament threads are zoomed and indicated with arrows. The last panel shows clear post-eruption arcades. 
			 \textit{Bottom row:} Reverse color soft X-ray images taken by the instrument XRT/Hinode. The filter wheel 1 is Open, while the filter wheel 2 is in the ``titanium/polyimide'' (Ti/poly) filter.
			 The field of view of all images is $300^{\prime\prime}$$\times $$300^{\prime\prime}$. The dates are shown as YYYY-MM-DD in all panels.}
   \label{2011_remote}
   \end{figure}

These eruptions originated from a polarity inversion line that ran between active regions 11226 and 11227 and into active region 11227.
The evolution of the photospheric line-of-sight (LOS) magnetic field of the two active regions is shown in Figure \ref{2011_remote}, top row, from 31 May to 3 June. Both ARs were characterized by a positive leading polarity. AR 11226 had a bipolar configuration, while AR 11227 encompassed two bipolar groups.

Table \ref{tbl:table} (first column) lists the sign of the magnetic helicity for this event as determined using the proxies and methods described in Section \ref{subs:magneticprior}. 
When the active regions were visible in the HMI field-of-view at a sufficiently large distance from the eastern limb (a few tens of degrees), they were already at an advanced stage of evolution, \textit{i.e.} no significant new flux emergence was observed. This means that it was not possible to infer the chirality of the active regions from the observation of magnetic tongues.

\begin{table}
\caption{}{Helicity proxies for the two events under analysis. ``H$\alpha$ filament'' refers to all the proxies visible in H$\alpha$ related to the chirality of a filament, \textit{e.g.} orientation of filament legs, barbs, and fibrils. ``EUV filament'' refers to the crossings of absorption and emission filament threads visible in EUV. Event 1 refers to the CME that occurred on 2 June 2011 and Event 2 refers to the CME on 14 June 2012.}\label{tbl:table}
\begin{center}
\begin{tabular}{lccc}     
\hline
\textbf{Proxy} & \textbf{Event 1} & \textbf{Event 2} \\
\hline
Magnetic tongues & - & Positive \\
H$\alpha$ filament &  -   & -\\
EUV filament  & Positive  & - \\
Filament Rotation & Positive & - \\
Sigmoid & Positive & Positive \\
Skew of overlying loops & Positive  & Positive \\
Flare ribbons &  - & - \\
\hline
\end{tabular}
\end{center}
\end{table}

The H$\alpha$ data (Figure \ref{2011_remote}, second row) show clearly the presence of a filament running along the PIL between the leading positive polarity and the adjacent negative polarity of AR 11227 and continuing in the region between active regions 11227 and 11226, forming a U-shaped structure. This is visible in Figure \ref{2011_remote}e, where the H$\alpha$ figure has been overlaid with magnetogram contours. The eastern leg of the U-shaped structure started to erupt at around 06:25 UT (Eruption 1), and by around 06:50 UT the filament had partially reformed. Eruption 2 began around 07:15 UT, and by 08:00 UT the filament had completely disappeared. Subsequent data (not shown) reveal that the filament had completely reformed by 15:00 UT.

In addition to being visible in H$\alpha$, the filament was also visible in the EUV 171 \AA \ waveband (Figure \ref{2011_remote}, third row), where it appeared as a combination of absorption (dark) and emission (bright) threads. Hence, we can discern overlying and underlying dark and bright threads and determine the chirality of the filament through the geometry of these crossings. All the crossings displayed in the figure suggest that the erupting filament had the positive helicity, knowing that the magnetic field is pointing roughly westward (see the axial field determination later in this section). In addition, the western section of the filament showed a clockwise rotation at around 07:40 UT (not shown in Figure \ref{2011_remote}), which is again a sign of right-handed chirality. However, the filament was thin and relatively short and hence we could not estimate the chirality using the characteristics of its barbs or legs and fibrils. 

Soft X-ray observations (Figure \ref{2011_remote}, bottom row) show a particularly strong emission coming from AR 11226 and from a few loops connecting AR 11227 and AR 11226. The connecting loops formed initially a double J-shaped structure (Figure \ref{2011_remote}m) that evolved later on 2 June into a continuous sigmoid (Figure \ref{2011_remote}n). The forward S-shape of the sigmoid is a sign of positive chirality. During both eruptions, however, only the eastern part of the sigmoid could be seen to erupt.  

Observations of the coronal arcades from both soft X-rays and EUV 171 \AA \, (Figures \ref{2011_remote}i, j, and n) show that, when viewed from the positive polarity side, the arcade above the PIL or filament channel was skewed to the right, implying the presence of a right-skewed arcade. This is again a sign of positive magnetic helicity.

Finally, we try to estimate the chirality from flare ribbons. As seen in H$\alpha$ (Figure \ref{2011_remote}, second row) and UV 1600 \AA \, (not shown) observations, both eruptions were associated with flare ribbons. However, the complexity of the flare ribbon structure makes the interpretation of their shape difficult and we do not consider them here for a chirality determination. 

Hence, we can conclude that all the used helicity proxies suggest that the erupting flux rope had a positive chirality. As the source region was in the southern hemisphere, this flux rope followed the hemispheric helicity rule (see Section \ref{subs:chirality}). 

\begin{figure}
   \centerline{
               \includegraphics[width=0.45\textwidth,clip=]{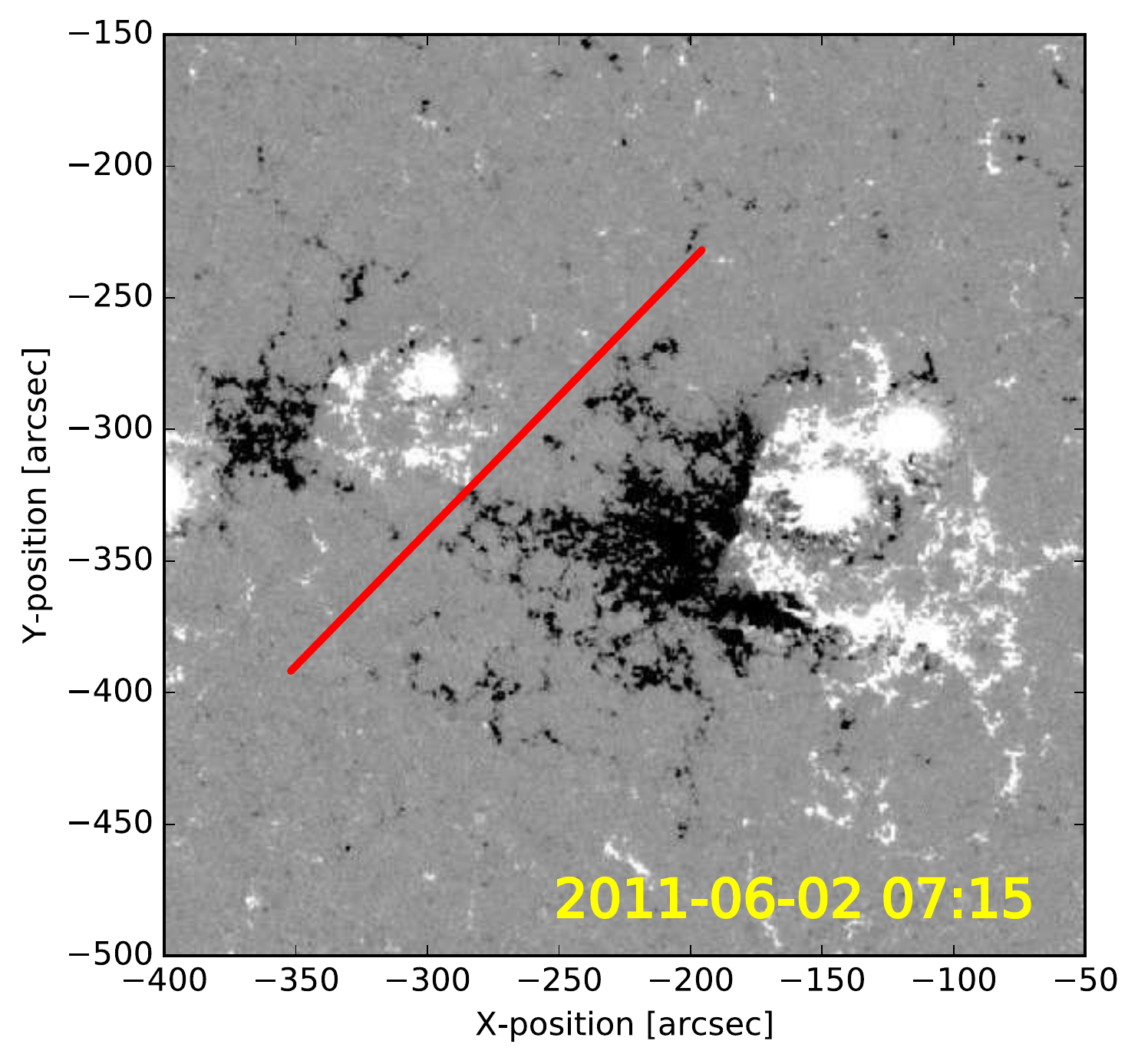}
               \includegraphics[width=0.45\textwidth,clip=]{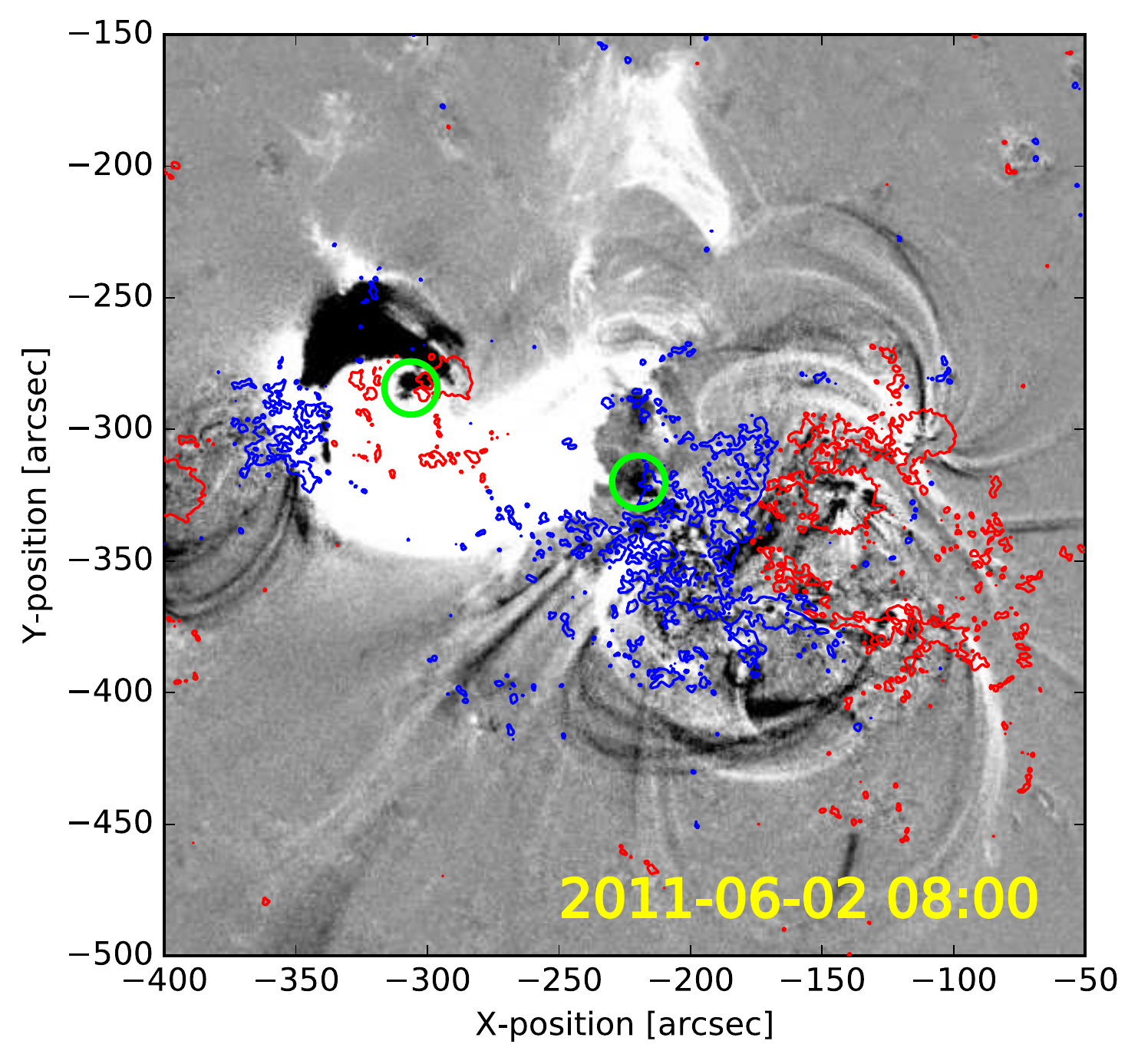}
              }
              \caption{\textit{Left:} HMI magnetogram showing the PIL approximated as a straight line (in red). \textit{Right:} Base-difference image of the region in 131 \AA \, saturated to $\pm 70$ DN s$^{-1}$ pixel$^{-1}$ overlaid with HMI magnetogram contours saturated to $\pm 200$ G (blue is used for the negative polarity and red for the positive polarity). The difference has been taken between the images at 08:00 UT (after the second eruption) and at 07:00 UT (between the two eruptions) on 2 June. The dimming regions (indicators of the flux rope footpoints) have been circled in green. The field of view of all images is $350^{\prime\prime}$$\times $$350^{\prime\prime}$. The dates are shown as YYYY-MM-DD in all panels.}
   \label{2011_tiltaxis} 
   \end{figure}

From SDO magnetograms (Figure \ref{2011_tiltaxis}, left panel), we infer the tilt of the PIL to be $|\tau| \simeq 45^{\circ}$ with respect to the ecliptic.  This is the dividing angle between bipolar and unipolar flux ropes, and hence four flux rope types with positive helicity are possible for this eruption: south-west-north (SWN), north-east-south (NES), west-north-east (WNE), and east-south-west (ESW). The post-eruption arcades for this event were very short and we could not use them to estimate the tilt of the axis. 

For a right-handed chirality magnetic field, the transverse magnetic field along the PIL is expected to point to the left when looking from the positive polarity side. The configuration of the photospheric magnetic polarities would then result in an axial field that is directed towards the north-west, \textit{i.e.} the flux rope would be of a SWN- or a WNE-type. In order to confirm the axial field prediction, we look at base-difference images at the EUV wavelength 131 \AA \, (Figure \ref{2011_tiltaxis}, right panel). As seen from the EUV dimmings, the western footpoint is rooted in the negative polarity region, and the eastern footpoint in the positive one. This means that the footpoints of the flux rope indicate that the flux rope axial field is indeed pointing to the west. In addition, the configuration of the filament from H$\alpha$ observations shows that the filament involved in Eruption 2 is directed from south\,--\,east to north\,--\,west. This confirms that we can expect either a SWN- or a WNE-type flux rope in interplanetary space. 
   
\subsubsection{In situ Observations}

\begin{figure}  
	\centerline{
	\includegraphics[width=0.95\textwidth,clip=]{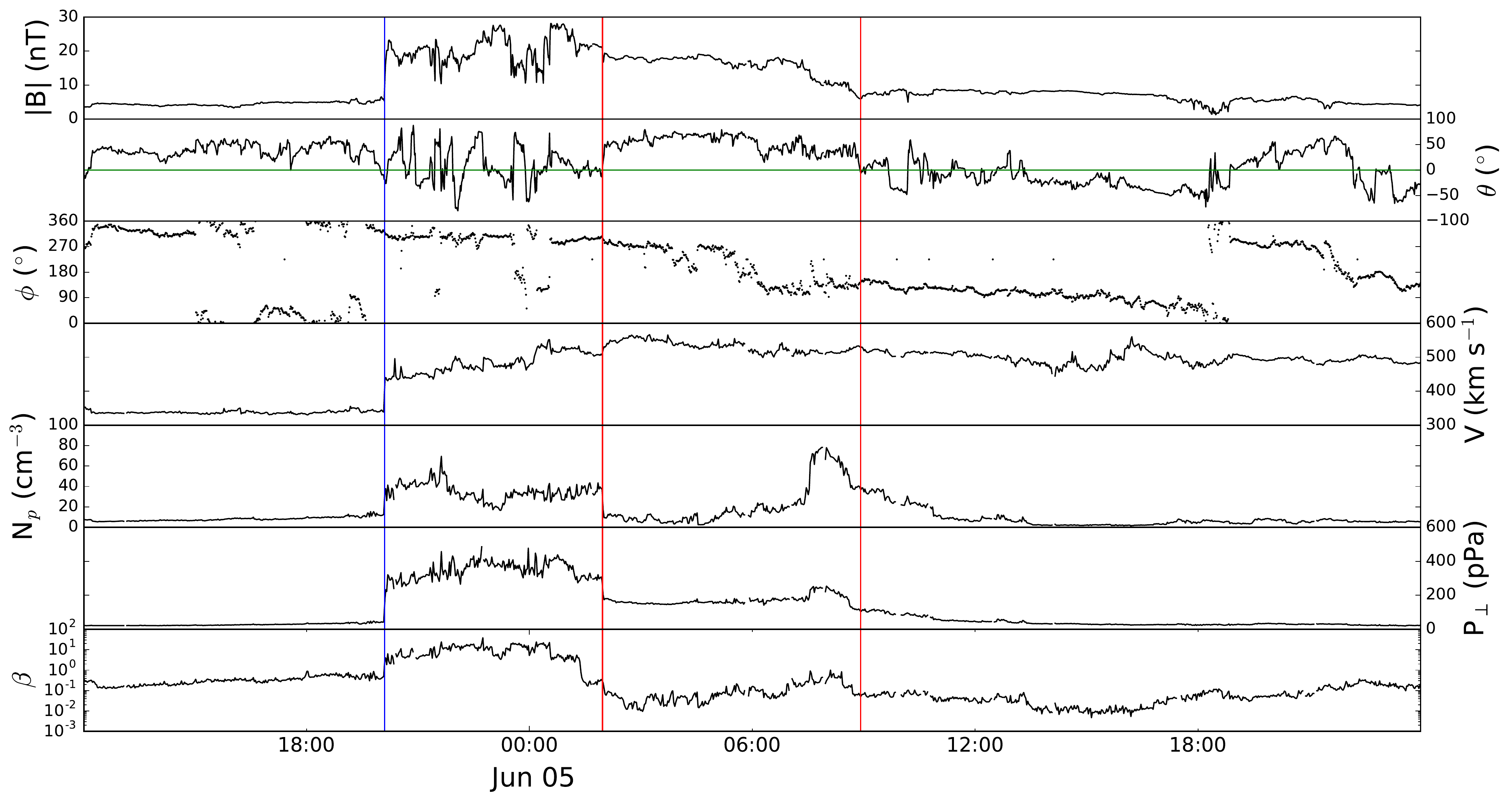}}
	\centerline{
	\includegraphics[width=0.45\textwidth,clip=]{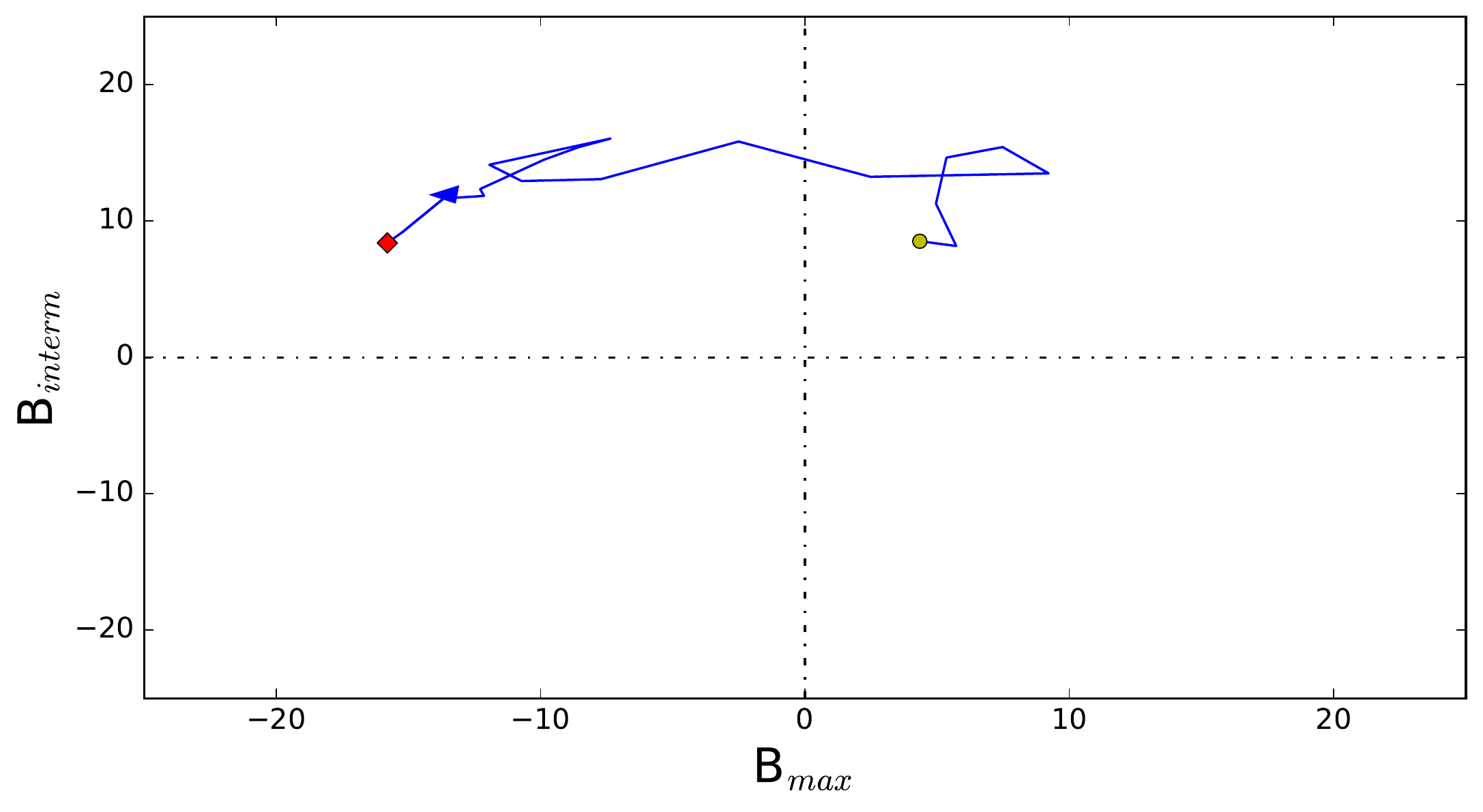}
	\includegraphics[width=0.45\textwidth,clip=]{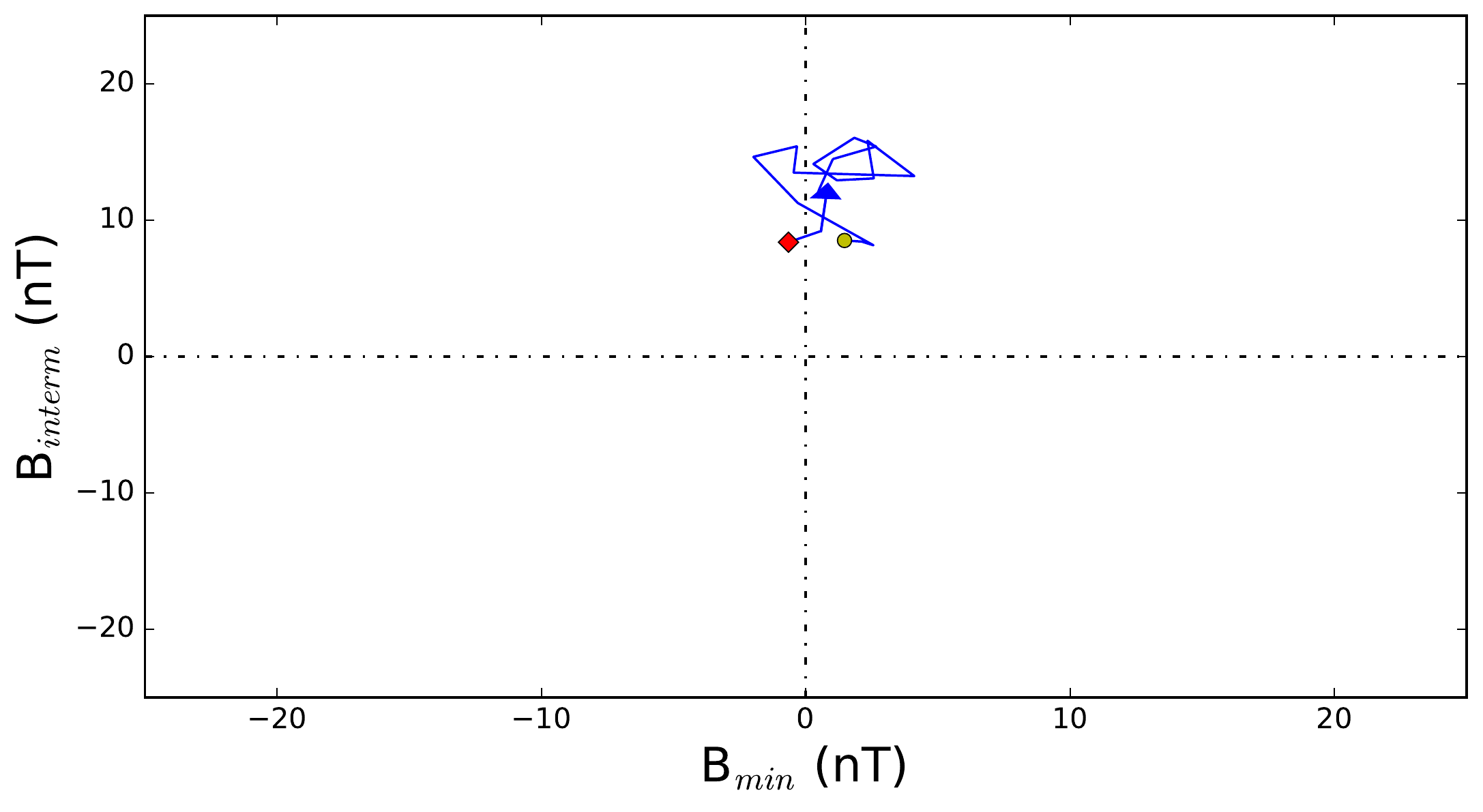}
	}
	\caption{\textit{Top:} The June 2011 CME as observed \textit{in situ} by \textit{Wind}. The blue line indicates the interplanetary shock, while the red lines indicate the leading and trailing edges of the flux rope. The parameters shown from top to bottom are: (a) magnetic field magnitude, (b) $\theta$ and (c) $\phi$ components in GSE angular coordinates, (d) solar wind speed, (e) proton density, (f) perpendicular pressure, and (g) plasma $\beta$. \textit{Bottom:} Results of the MVA for the June 2011 CME, showing the rotation of the magnetic field vectors in the $B_{max}$-$B_{interm}$ plane (\textit{left}) and in the $B_{min}$-$B_{interm}$ plane (\textit{right}). The start of the rotation is indicated by the red diamond, the direction of the rotation by the arrow, and the end point by the yellow dot. The magnetic field data have been interpolated to a 20-minute cadence.}
	\label{2011_insitu}
\end{figure}

The interplanetary shock associated with the CME was detected by \textit{Wind} on 4 June 2011 at 20:06 UT. Flux rope signatures could be identified on 5 June from $\sim$02:00 UT to $\sim$09:00 UT (Figure \ref{2011_insitu}, top panel), \textit{i.e.} enhanced magnetic field combined with smooth rotation of the magnetic field direction and depressed plasma $\beta$ \citep[\textit{e.g.},][]{burlaga1981}. The plasma $\beta$ is the ratio of the plasma pressure to the magnetic pressure. The visual inspection of the magnetic field measurements shows that within the ICME the magnetic field rotates from west to east pointing strongly northward at the center. Hence, the flux rope has a WNE flux rope topology. This is also seen by noting that $\Delta\phi<0$ and that $\theta>0^{\circ}$, where $\theta$ and $\phi$ are the latitudinal and longitudinal components of the magnetic field, respectively. This means that our expectations of finding either a SWN- or a WNE-type flux rope are satisfied.

The results of the MVA are shown in Figure \ref{2011_insitu}, bottom panels. The ratio of the intermediate-to-minimum eigenvalues is $\lambda_{2}/\lambda_{3} = 3$, confirming the validity of the method. The rotation shown in the $B_{max}-B_{interm}$ plane corresponds to the one of a WNE-type flux rope. The orientation of the axis from the MVA is $(\theta_{A},\phi_{A})=(68^{\circ},139^{\circ})$, \textit{i.e.} consistent with a highly inclined flux rope. 

The angle between the shock normal and the radial direction is $\alpha = 33.0^{\circ}$, which indicates that the spacecraft cut the ICME quite far from its nose. The perpendicular pressure profile (Figure \ref{2011_insitu}f) shows a clear plateau-like profile and can therefore be associated to group 2 (see Section \ref{subs:magneticinsitu}). This means that the spacecraft crossed the ICME further from its central axis, perhaps at its outer edge. This is also consistent with the relatively weak magnetic field rotation as seen from the visual inspection of the magnetic field components and from the hodogram in Figure \ref{2011_insitu}.

The GSR also suggests that the ICME was encountered far from the axis. In such a case, the GSR results are not reliable \citep{isavnin2011} and we do not consider them here. 

\subsection{Event 2: CME on 14 June 2012}
\label{subs:june2012}

Our second case study was associated with a very well-defined flux rope observed by \textit{Wind} on 16\,--\,17 June 2012. We found that the corresponding CME erupted on 14 June from AR 11504. The same CME association has been made by Richardson and Cane in their Near-Earth Interplanetary Coronal Mass Ejections list\footnotemark[5] and by \citet{kubicka2016}. We again first perform the remote-sensing analysis and then proceed to \textit{in situ} observations. 

\footnotetext[5]{\url{http://www.srl.caltech.edu/ACE/ASC/DATA/level3/icmetable2.htm}}

\subsubsection{Coronal Observations}

A large symmetrical full halo CME erupted on 14 June. The event was first detected by LASCO C2 on 14 June at 14:12 UT, having a plane-of-sky linear speed of 987 km s$^{-1}$. The same event appeared in the STEREO A COR1 field-of-view emanating from the SE-quadrant on 14 June at 13:25 UT, and in the STEREO B COR1 field-of-view emanating from the SW-quadrant on 14 June at 13:45 UT. The CME appeared in the white-light images of both STEREO spacecraft as a classic three-part CME, \textit{i.e.} consisting of a bright front, cavity, and core.

\begin{figure}
   \vspace{0.05\textwidth}	
   
   \centerline{
               \includegraphics[width=0.27\textwidth,clip=]{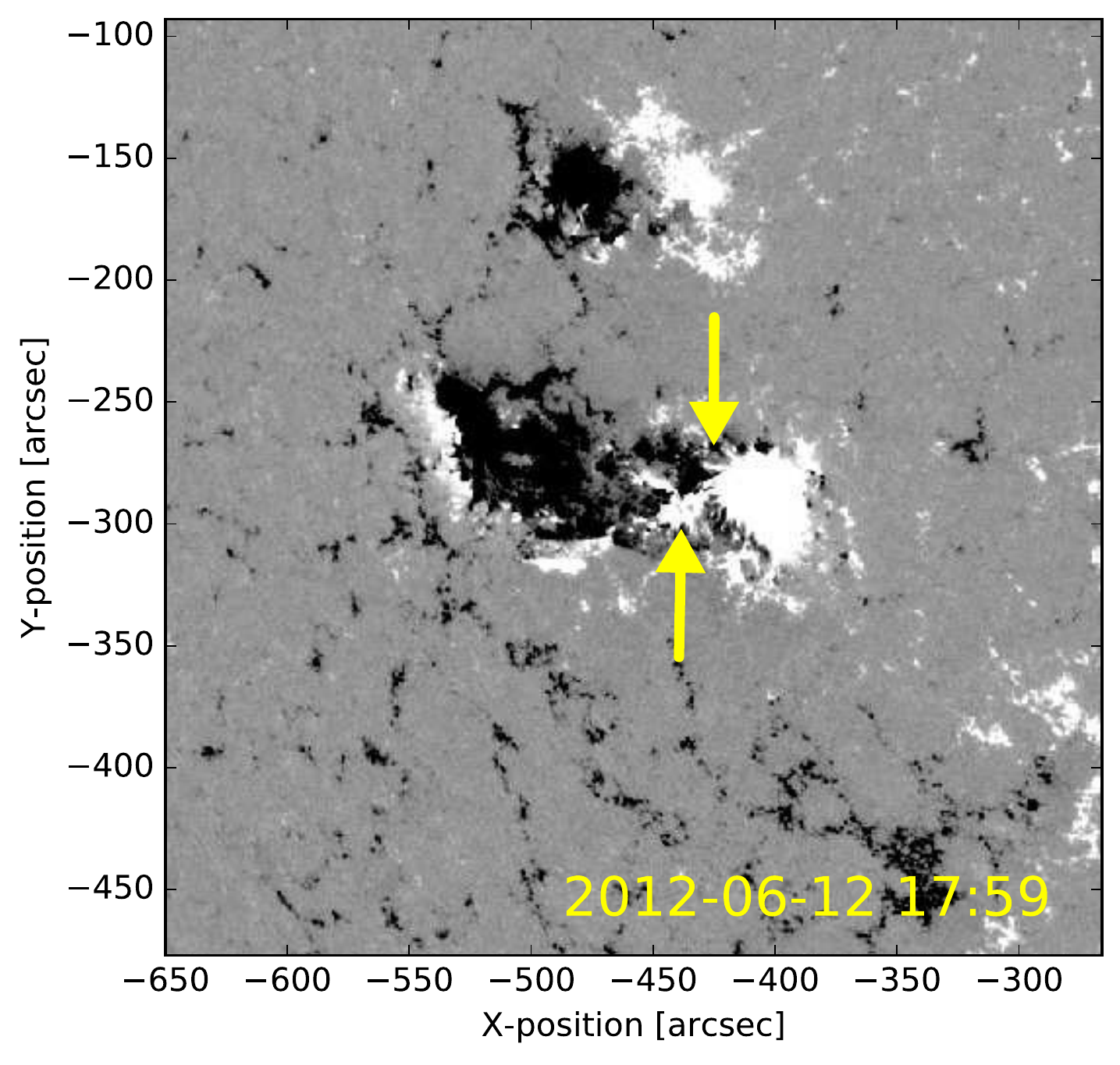}
               \includegraphics[width=0.235\textwidth,clip=]{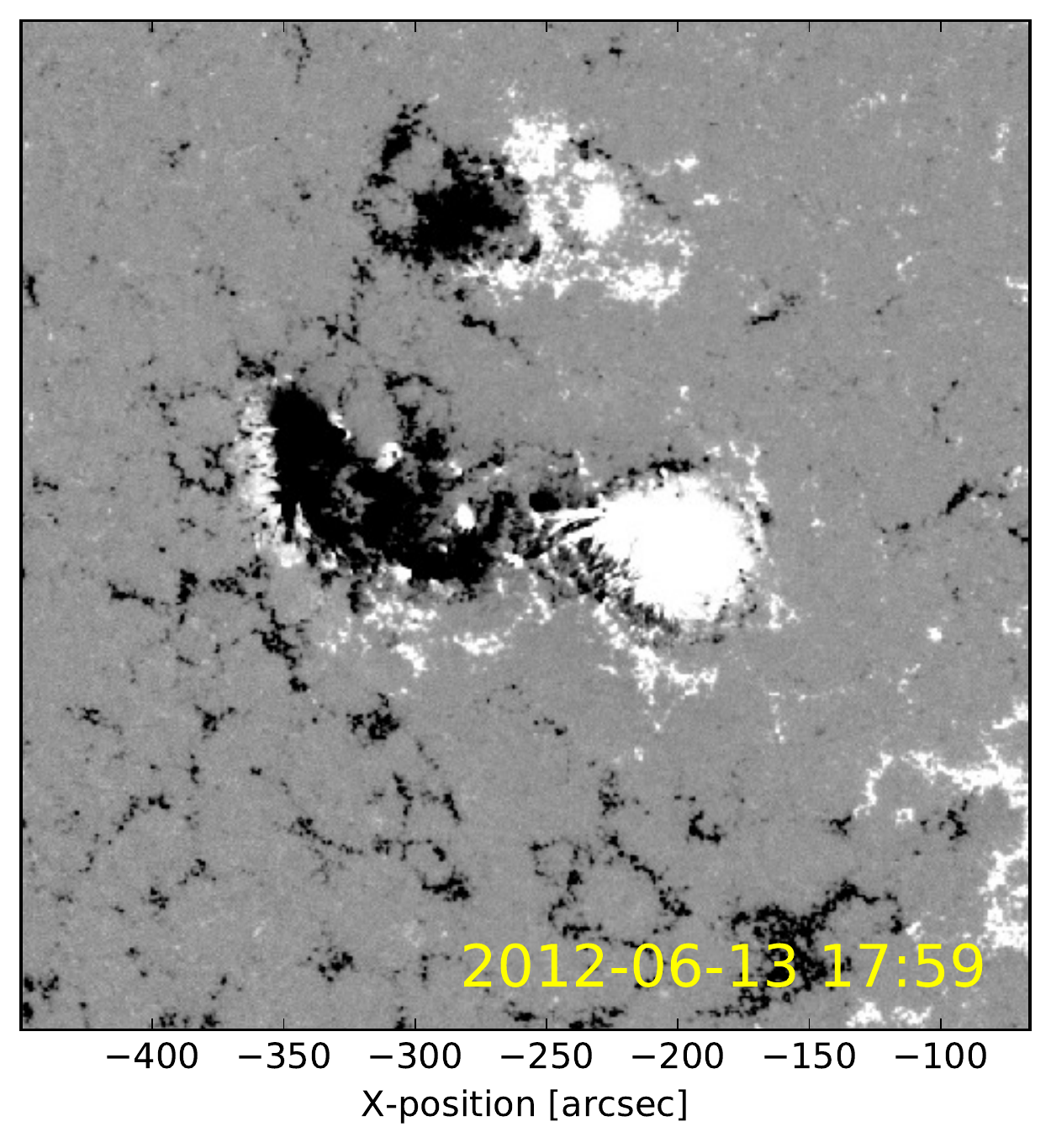}
			  \includegraphics[width=0.235\textwidth,clip=]{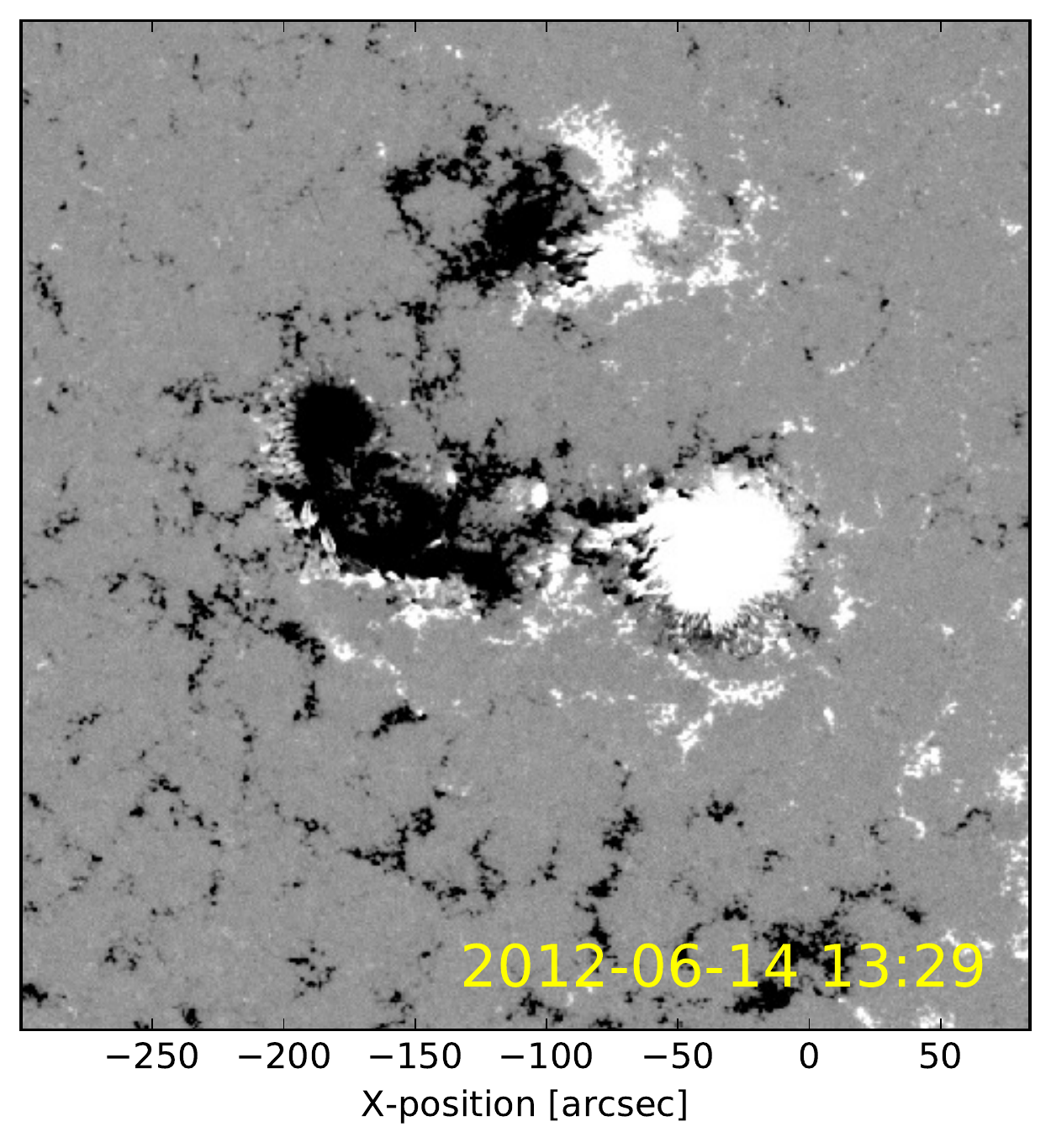}
               \includegraphics[width=0.235\textwidth,clip=]{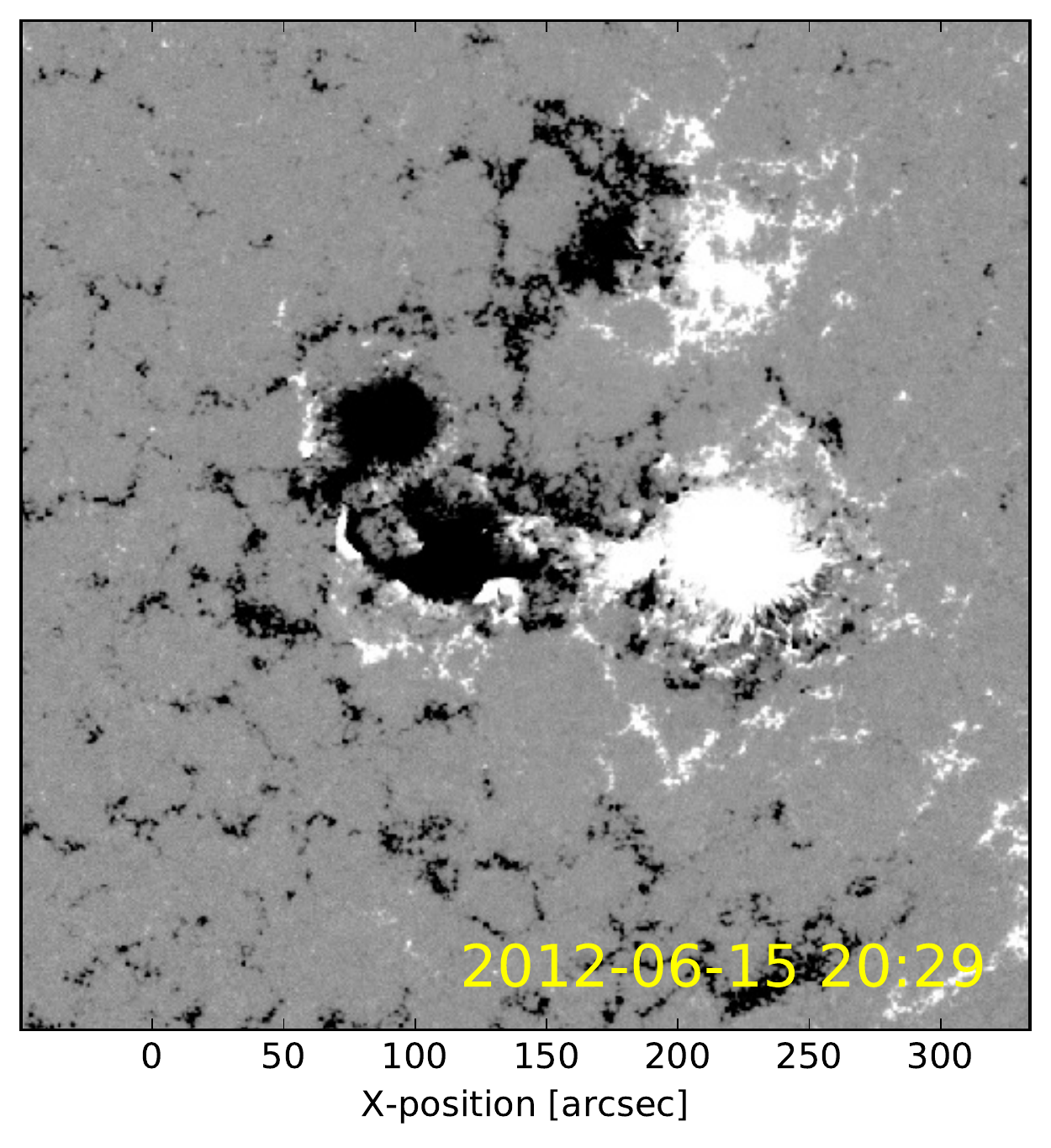}
			  }
			  
      \vspace{-0.31\textwidth}   
     \centerline{\large \bf                         
      \hspace{0.29 \textwidth} {HMI MAGNETOGRAM}
      	\hfill}
      	
      \vspace{0.017\textwidth}   
     \centerline{\normalsize \bf    
      \hspace{0.03 \textwidth}  \color{white}{(a)}
      \hspace{0.18 \textwidth}  \color{white}{(b)}
      \hspace{0.179 \textwidth}  \color{white}{(c)}
      \hspace{0.182\textwidth}  \color{white}{(d)}
         \hfill}
      	
     \vspace{0.30\textwidth}
 
 	\centerline{
               \includegraphics[width=0.27\textwidth,clip=]{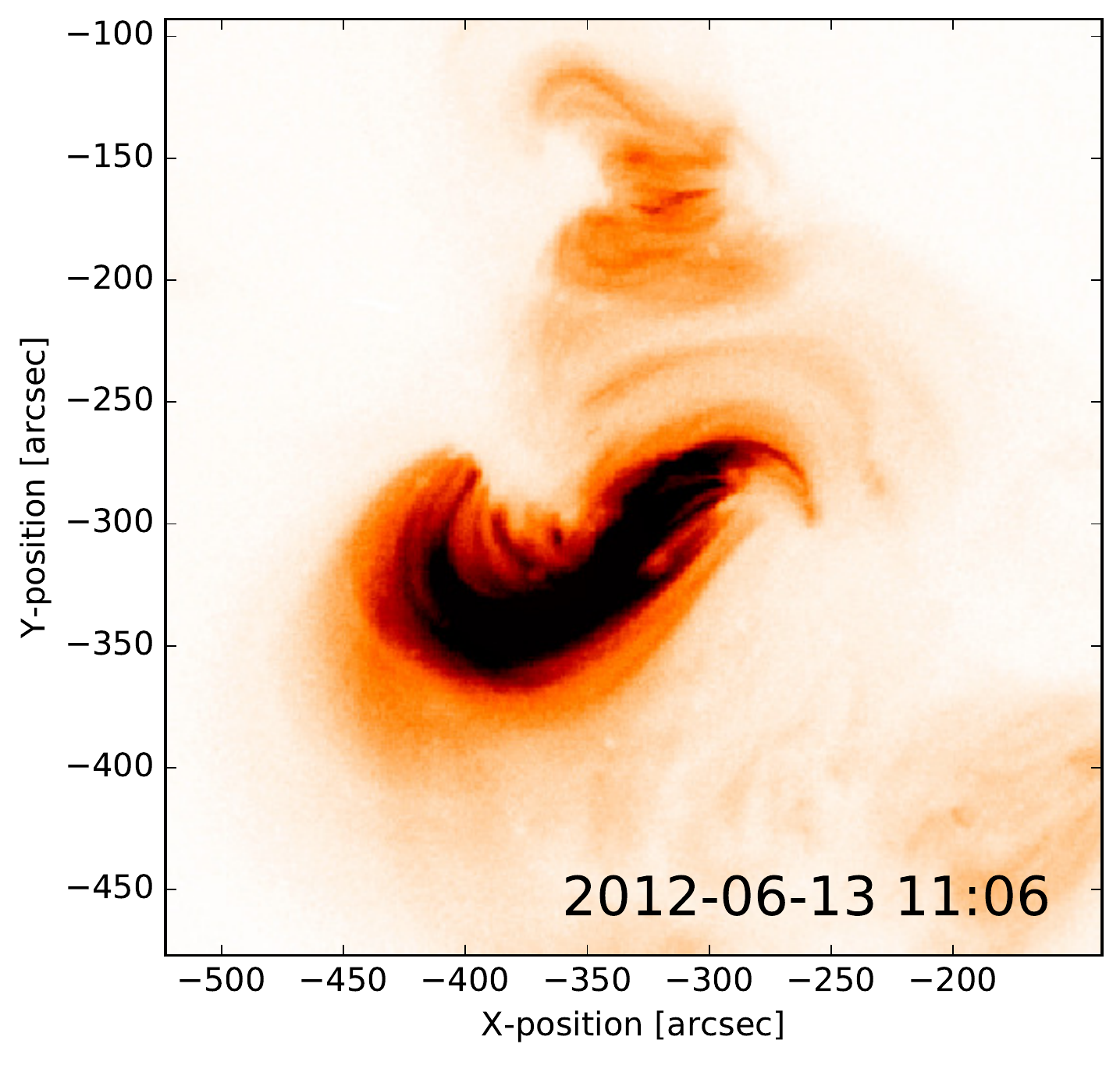}
               \includegraphics[width=0.235\textwidth,clip=]{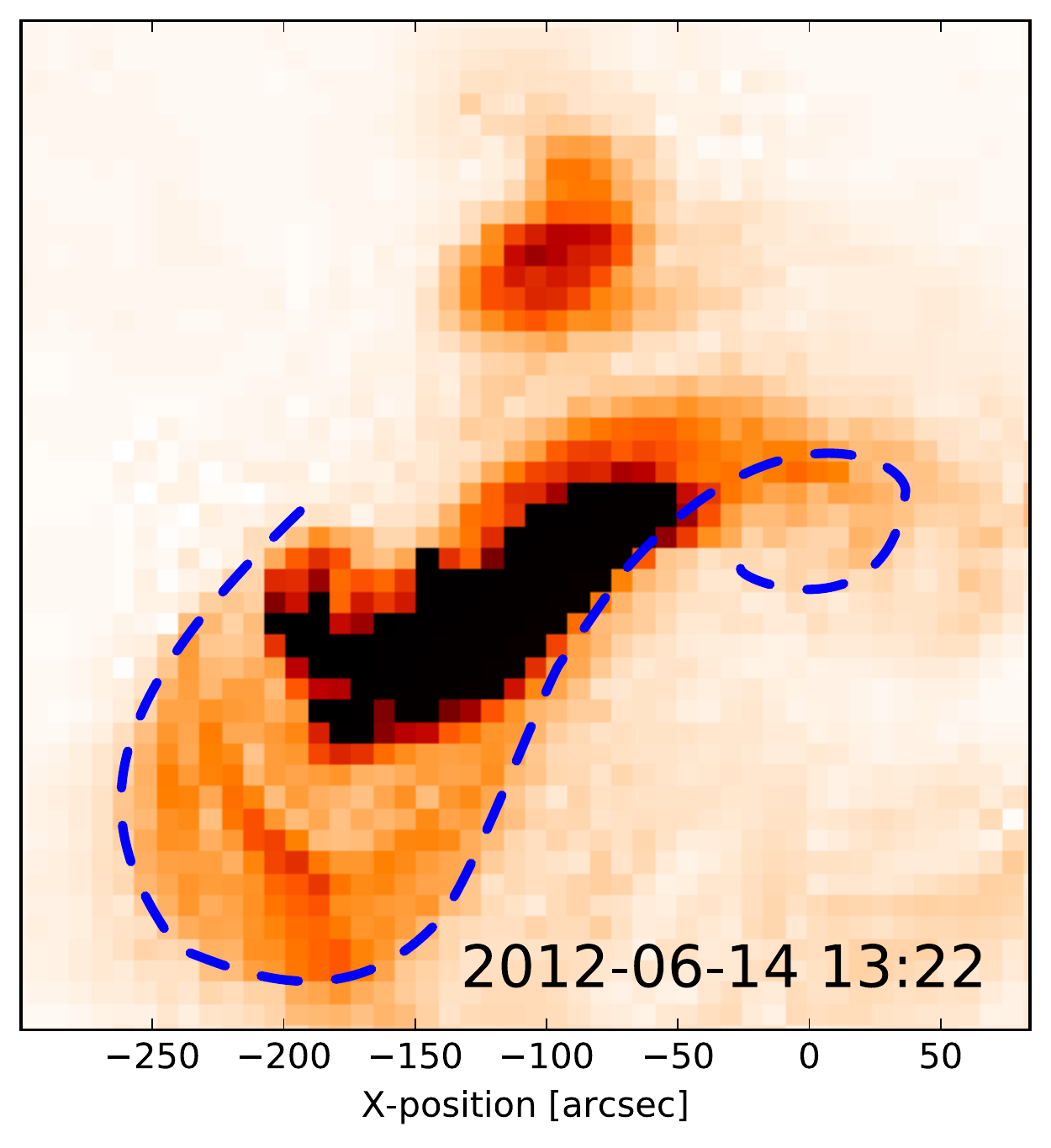}
			  \includegraphics[width=0.235\textwidth,clip=]{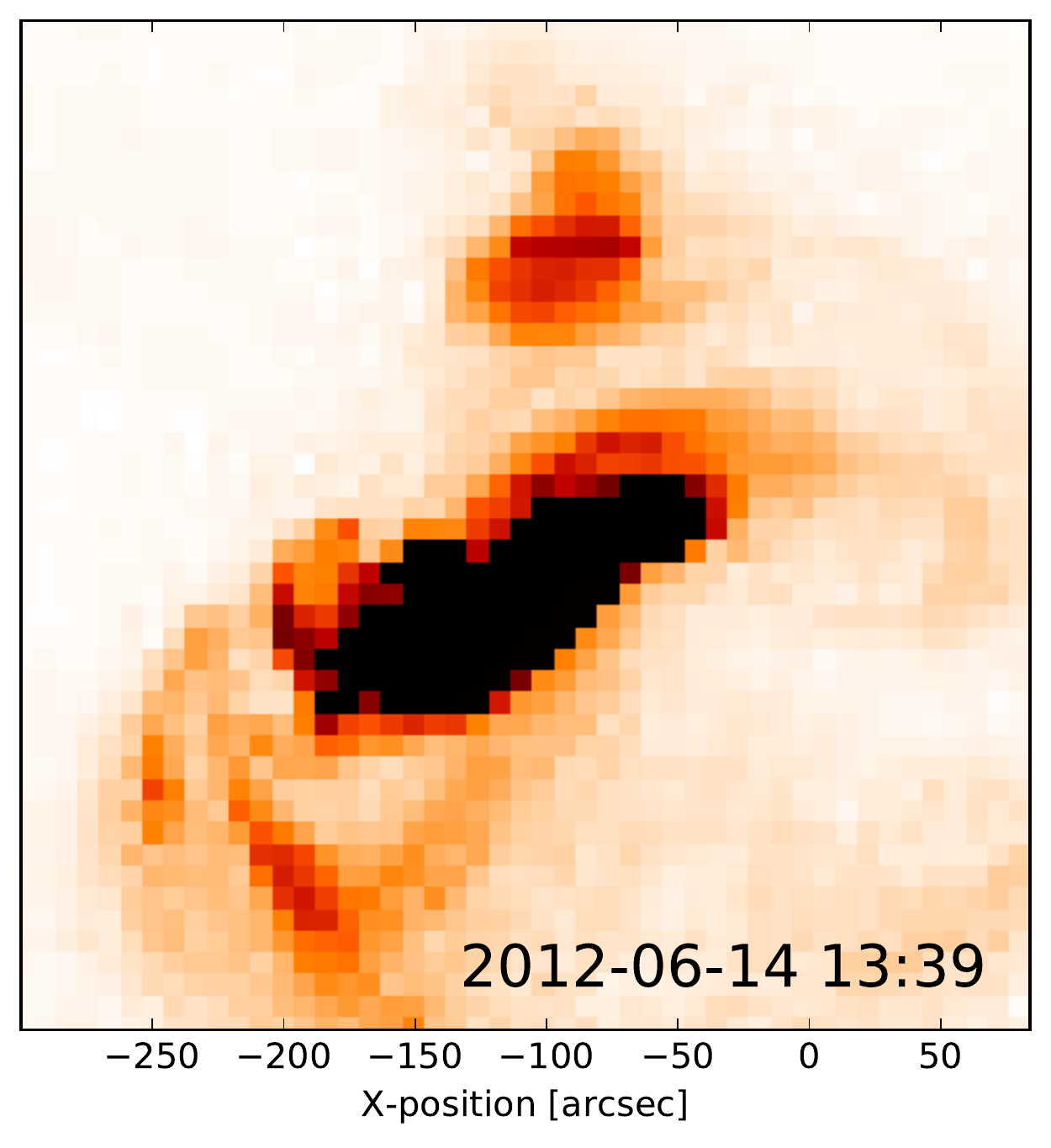}
               \includegraphics[width=0.235\textwidth,clip=]{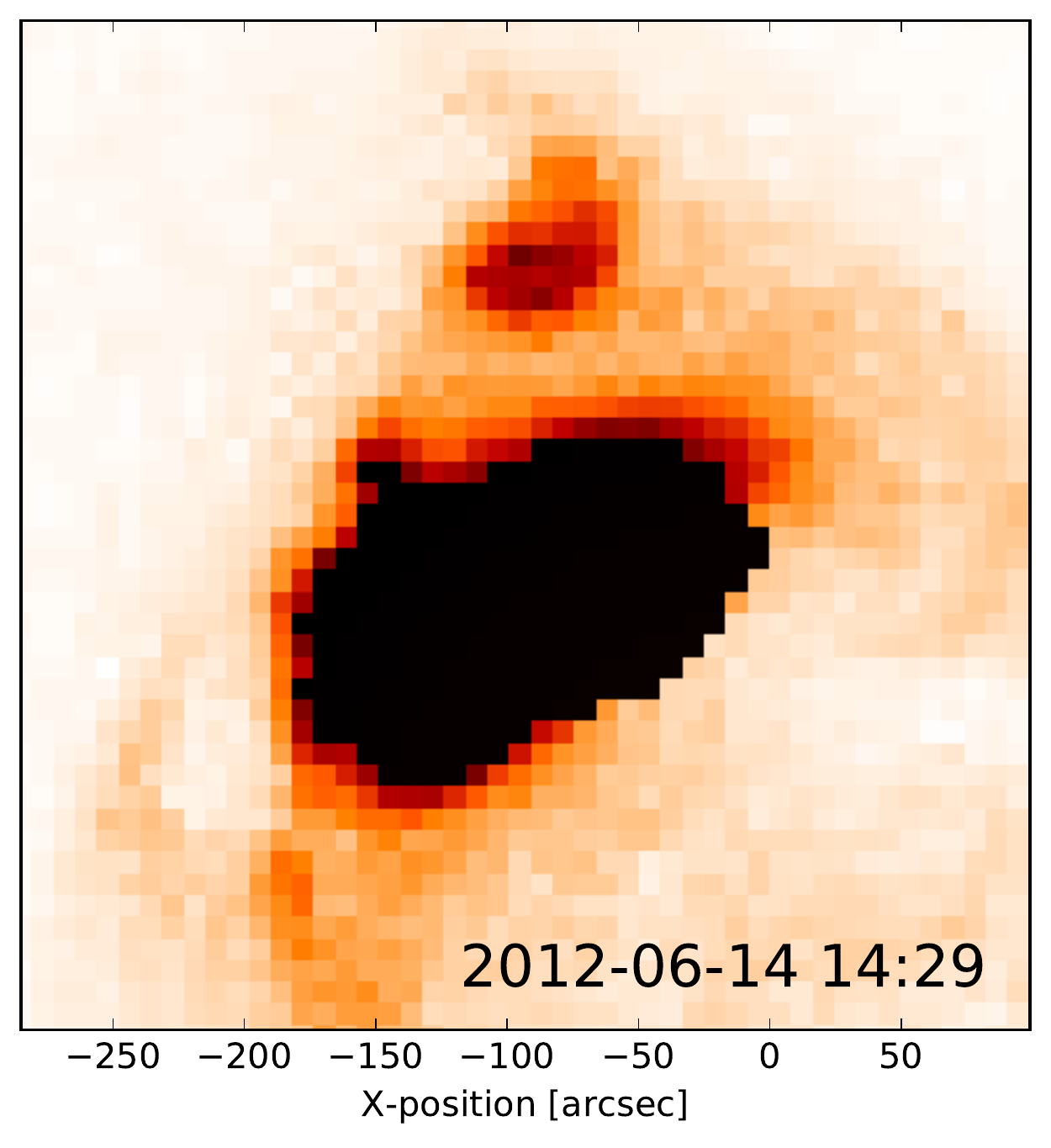}
			  }
              
      \vspace{-0.31\textwidth}   
     \centerline{\large \bf                         
      \hspace{0.33 \textwidth} {XRT SOFT X-RAYS}
      	\hfill}
      	
      	\vspace{0.017\textwidth}   
     \centerline{\normalsize \bf    
      \hspace{0.03 \textwidth}  \color{black}{(e)}
      \hspace{0.182 \textwidth}  \color{black}{(f)}
      \hspace{0.183 \textwidth}  \color{black}{(g)}
      \hspace{0.181\textwidth}  \color{black}{(h)}
         \hfill}
      	
     \vspace{0.3\textwidth}
     
     \centerline{
               \includegraphics[width=0.27\textwidth,clip=]{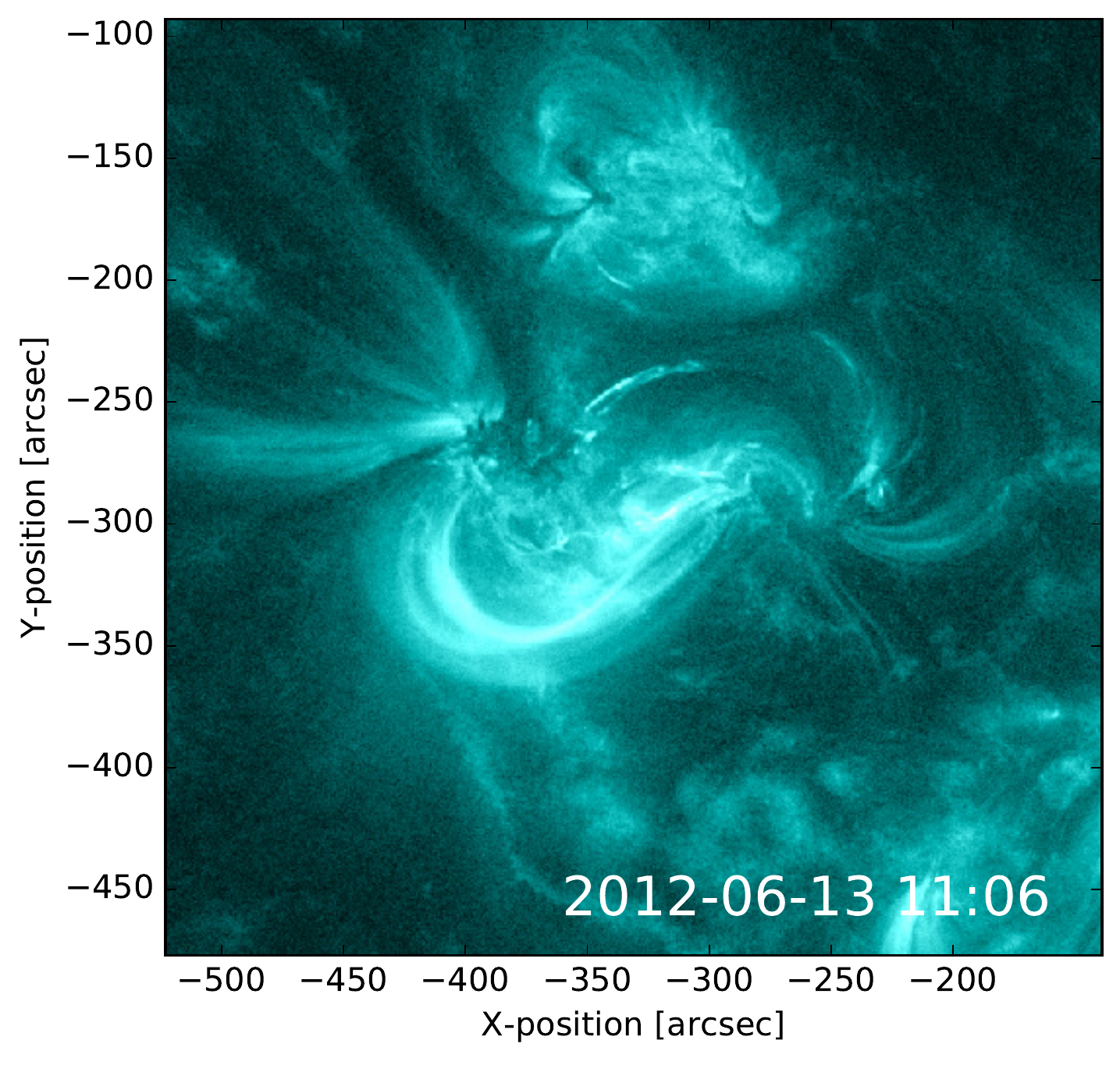}
               \includegraphics[width=0.235\textwidth,clip=]{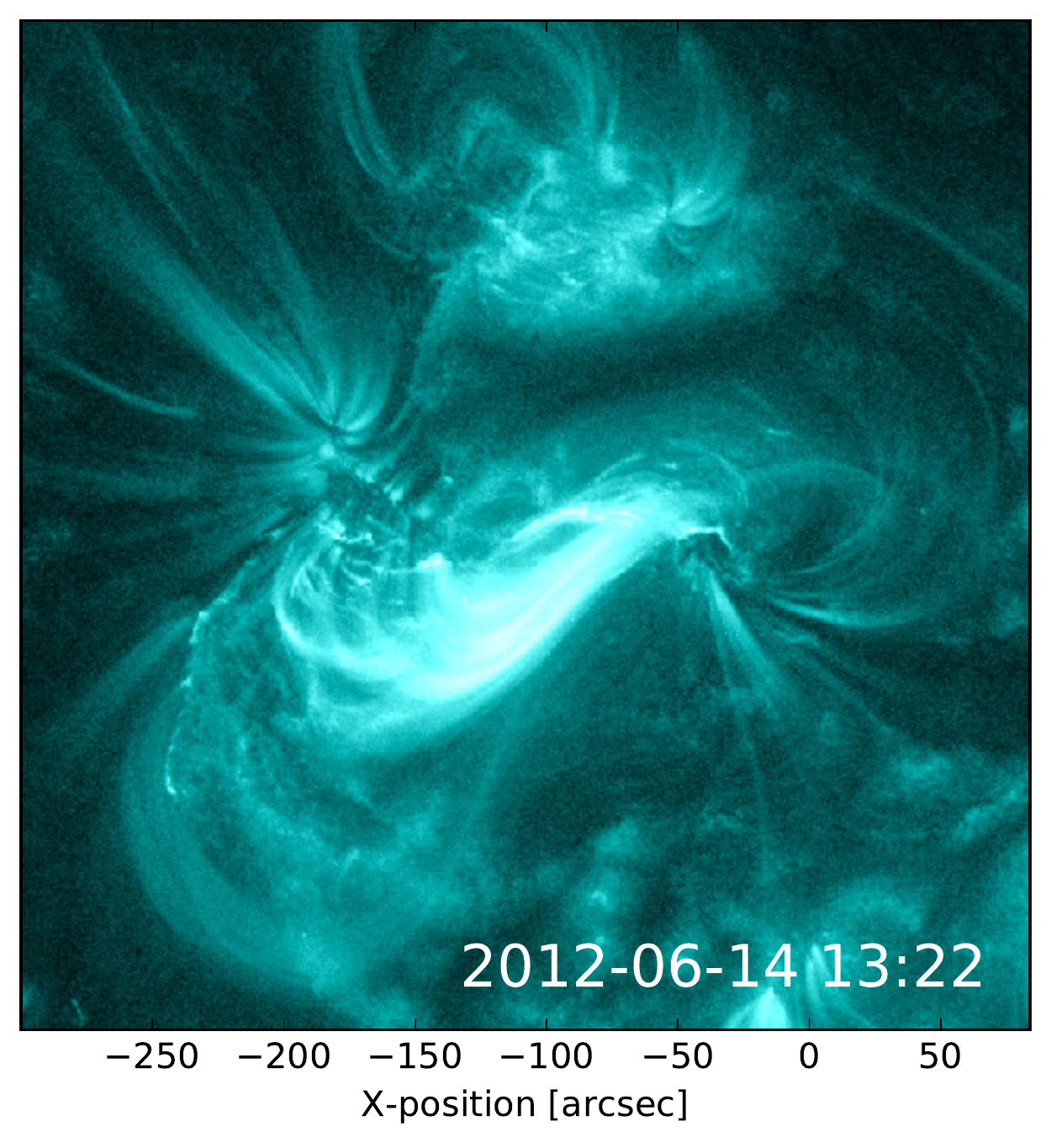}
			  \includegraphics[width=0.235\textwidth,clip=]{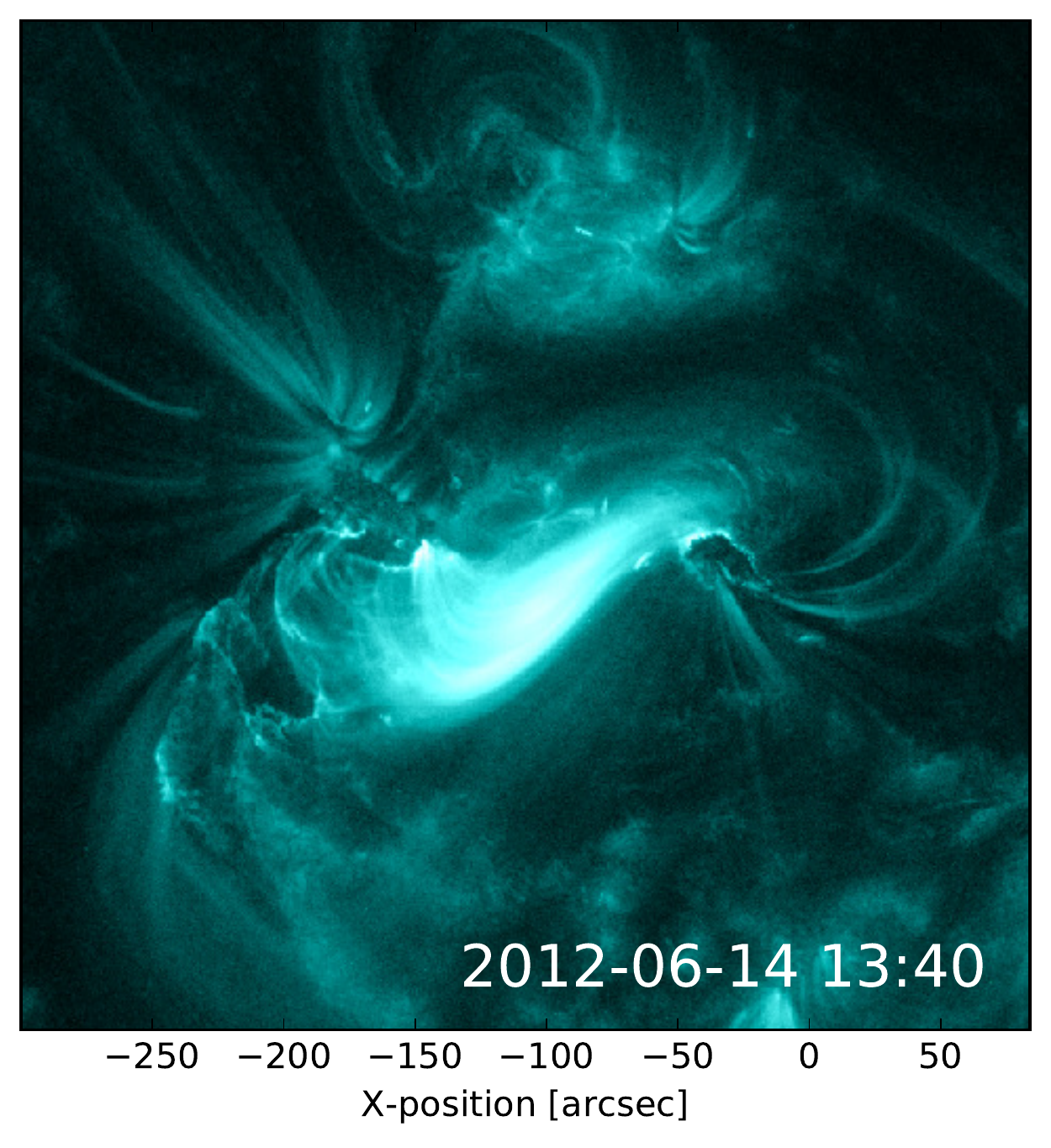}
               \includegraphics[width=0.235\textwidth,clip=]{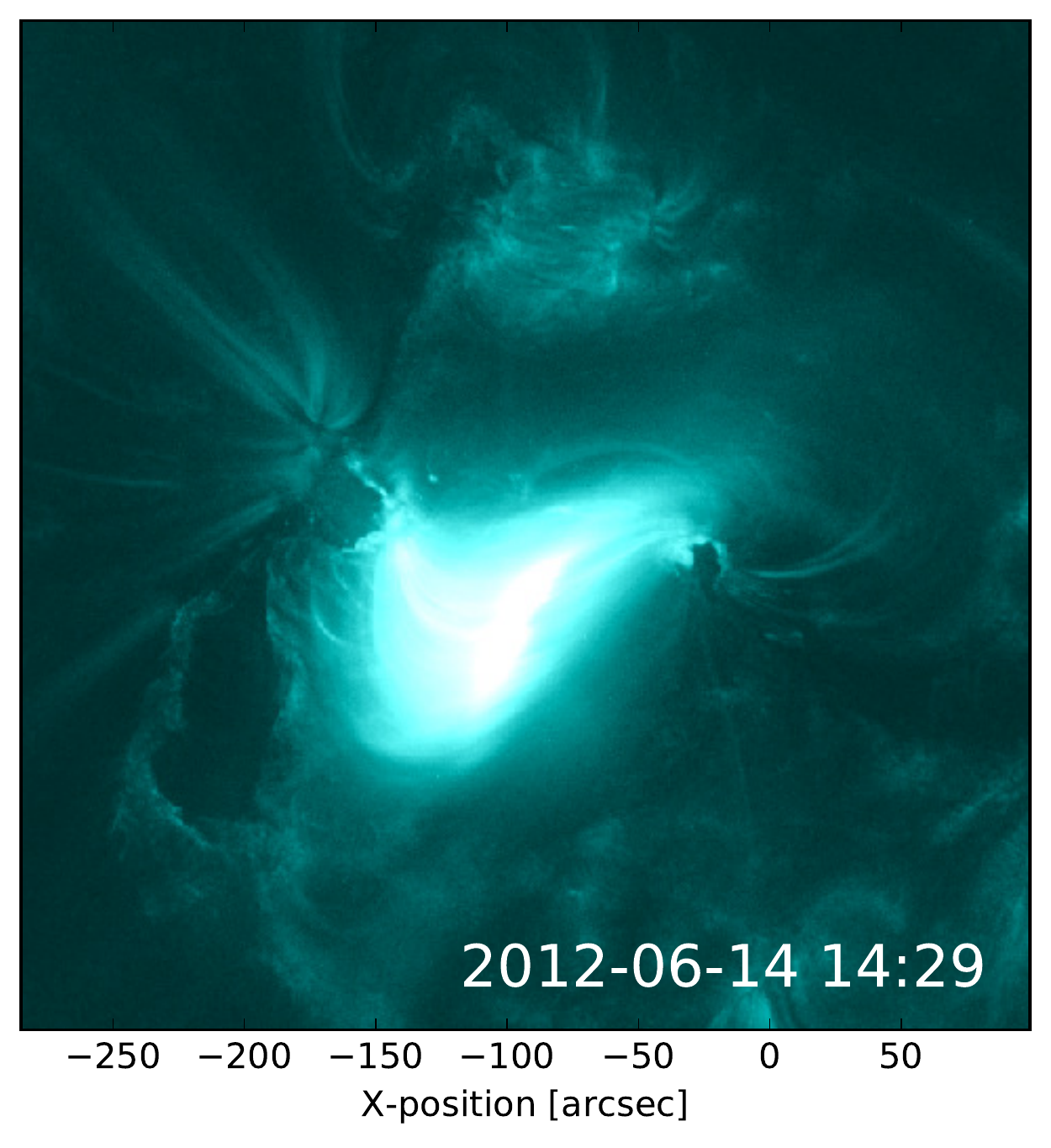}
			  }
              
      \vspace{-0.31\textwidth}   
     \centerline{\large \bf                         
      \hspace{0.36 \textwidth} {AIA EUV 131 \AA}
      	\hfill}
      	
      	\vspace{0.017\textwidth}   
     \centerline{\normalsize \bf    
      \hspace{0.03 \textwidth}  \color{white}{(i)}
      \hspace{0.187 \textwidth}  \color{white}{(j)}
      \hspace{0.187 \textwidth}  \color{white}{(k)}
      \hspace{0.18\textwidth}  \color{white}{(l)}
         \hfill}
      	
     \vspace{0.25\textwidth}
     
              \caption{\textit{Top row:} Evolution of AR 11504 in magnetograms as observed with HMI/SDO and saturated to $\pm 200$ G. The first image shows the presence of magnetic tongues (indicated by the yellow arrows), where the positive leading polarity extends under the negative trailing polarity.  \textit{Middle row:} Reverse colour soft X-ray images taken with XRT/\textit{Hinode}. The filter wheel 1 is Open, while the filter wheel 2 is in the ``titanium/polyimide'' (Ti/poly) filter. The shape of the sigmoid (from the EUV 131 \AA \, observations, bottom row) is outlined with the blue dashed line. \textit{Bottom row:} Coronal evolution in EUV of the eruptive event during 14 June, as observed with AIA/SDO. The images are taken with the 131 \AA \, filter. The field of view of all images is $384^{\prime\prime}$$\times $$384^{\prime\prime}$. The dates are shown as YYYY-MM-DD in all panels.}
   \label{2012_remote}
   \end{figure}

The photospheric magnetic field evolution of AR 11504 is shown in Figure \ref{2012_remote}, top row, from 12 June to 15 June. This active region appeared to be in its early stages based on the presence of new flux emergence, which means that magnetic tongues (if present) can be used as a proxy for chirality. Magnetic tongues were visible around 12 June at 18:00 UT. They show the leading positive polarity extending to the south of the trailing negative one, which indicates the presence of positive chirality. The summary of the used helicity sign proxies is shown in Table 1 (second column).

H$\alpha$ observations show that for this event no filament material was present at the PIL location. Hence, we cannot analyze the details of filament fine structures to determine the sign of the helicity. However, for this event we can get a proxy of the chirality using sigmoid and coronal arcade observations. Soft X-ray observations reveal a sheared arcade on 13 June at around 06:00 UT (Figure \ref{2012_remote}e). The coronal loop system appeared to be right-skewed, and was therefore right-handed relative to the PIL. A sigmoidal structure (Figure \ref{2012_remote}f) started to become faintly visible in soft X-rays at around 11:20 UT on 14 June.  
The forward S-shaped structure of the sigmoid is a sign of positive magnetic helicity. Unfortunately, high-resolution soft X-ray data are not available during 14 June. Therefore, in order to support the X-ray observations we observe the sigmoid also at the EUV 131 \AA\, waveband. Figure \ref{2012_remote}, bottom row, shows the evolution of the erupting structure at 131\,\AA, that follows closely the soft X-ray emission. 

Indications of the CME eruption started to be visible on 14 June at around 13:30 UT. The eruption was mostly detectable through its coronal dimmings, visible in difference EUV images, and the formation of a flare arcade. The CME was first detected in white light with STEREO A COR1 already at 13:25 UT, suggesting that the eruption took place high up in the corona. For a detailed description of the eruption and the formation of the eruptive structure see \citet{james2017}. 

Hence, we can conclude that all proxies that we applied suggest that the erupting flux rope had a positive chirality. Therefore, also this event followed the hemispheric rule for the helicity sign (see Section \ref{subs:chirality}). 

\begin{figure}
   \centerline{\hspace*{0.015\textwidth}
               \includegraphics[width=0.45\textwidth,clip=]{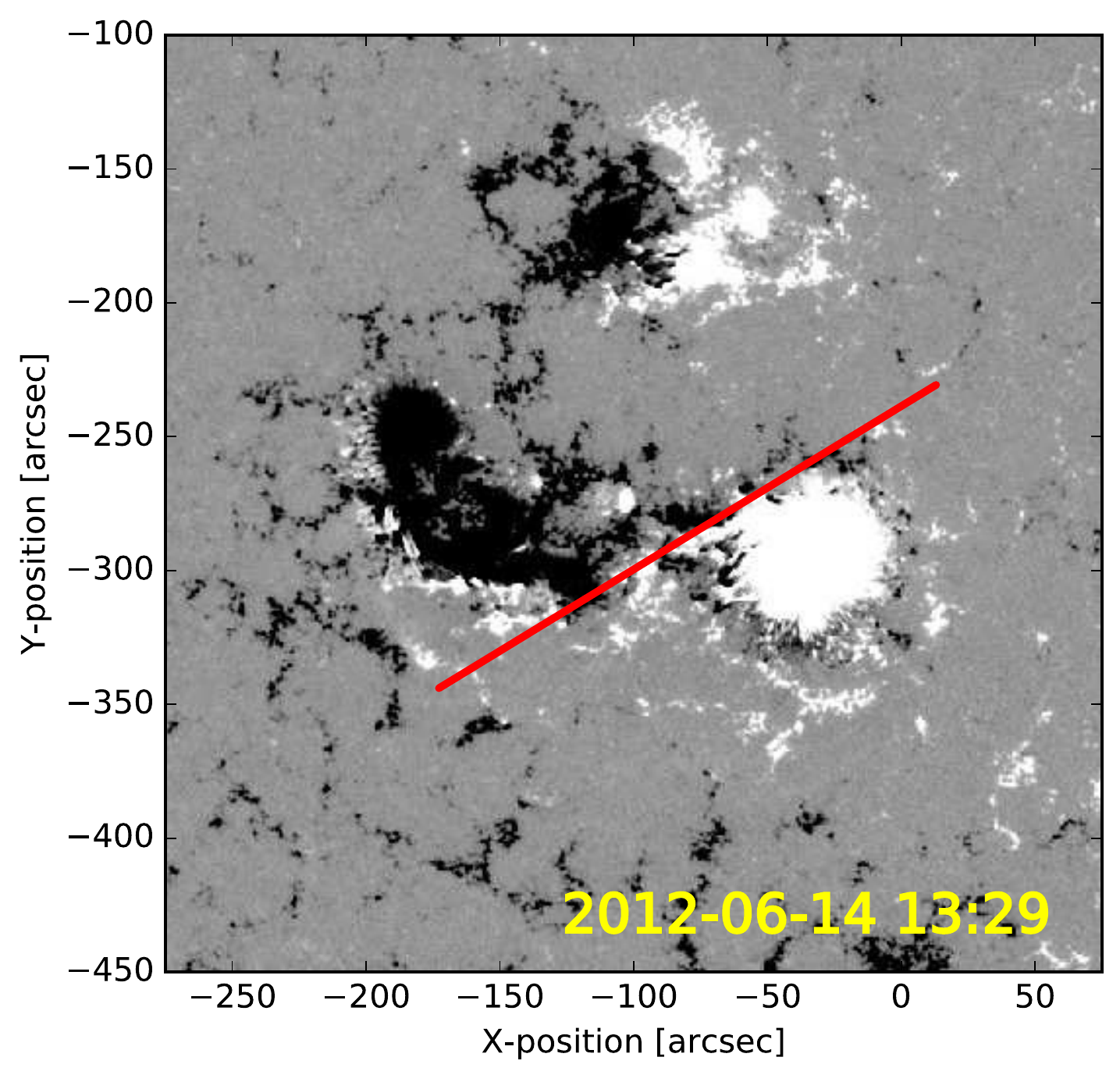}         
               \includegraphics[width=0.45\textwidth,clip=]{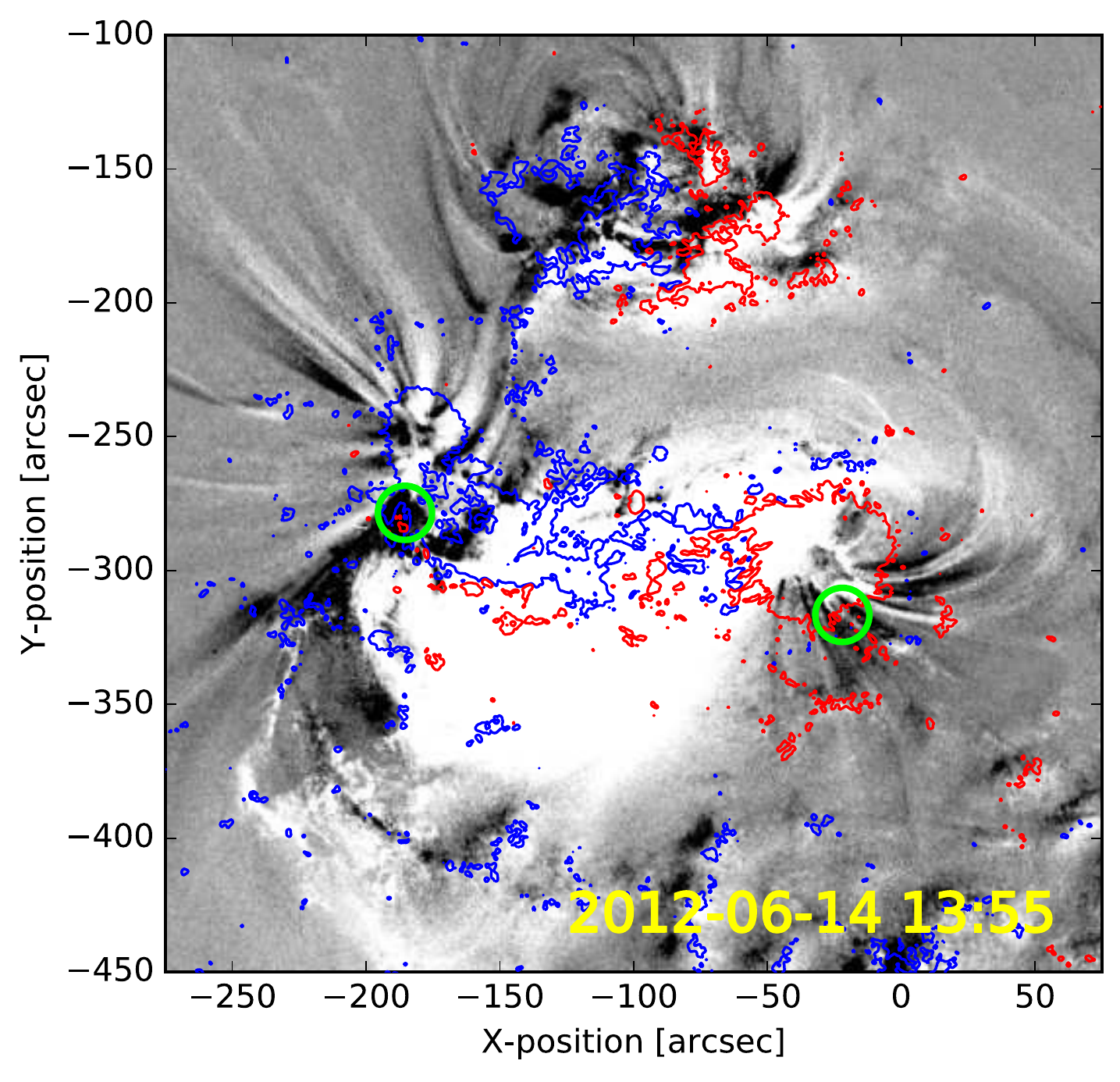}
              }
              \caption{\textit{Left:} HMI magnetogram showing the PIL approximated as a straight line (in red). \textit{Right:} Base-difference image of the region in 131 \AA \, saturated to $\pm 70$ DN s$^{-1}$ pixel$^{-1}$ overlaid with HMI magnetogram contours saturated to $\pm 200$ G (blue is used for the negative polarity and red for the positive polarity). The difference has been taken between the images at 13:55 UT and at 10:30 UT on 14 June. The dimming regions (indicators of the flux rope footpoints) have been circled in green. For a detailed description of the footpoints determination see \citealp{james2017}. The field of view of all images is $350^{\prime\prime}$$\times $$350^{\prime\prime}$. The dates are shown as YYYY-MM-DD in all panels.}
   \label{2012_tiltaxis} 
   \end{figure}

From SDO magnetograms (Figure \ref{2012_tiltaxis}, left panel), we infer the tilt of the PIL to be at an angle $|\tau| \simeq 30^{\circ}$ with respect to the ecliptic. Hence, we can expect a bipolar flux rope at 1 AU with positive helicity, \textit{i.e.} the possible flux rope types are: south-west-north (SWN) and north-east-south (NES). Again, the post-eruptive arcades were short and not well-defined, so we could not use them for inferring the tilt of the flux rope. 

For a right-handed region, the transverse magnetic field along the PIL is expected to point leftward when looking from the positive polarity side. The magnetic configuration of AR 11504 results then in an eastward axial field, which implies the erupting flux rope to be of NES-type. In order to confirm the axial field prediction, we look again at base-difference images at the EUV wavelength 131 \AA \, (Figure \ref{2012_tiltaxis}, right panel). As seen from the EUV dimmings, the western footpoint is rooted in the positive polarity region, and the eastern footpoint in the negative one. This means that the footpoints are aligned roughly eastward, suggesting that the axial field of the flux rope is also eastward. The right-handedness of the region implies that the leading field of an eastern-axial flux rope is oriented northward, and the trailing field is southward. This confirms our earlier suggestion of a NES-type flux rope prior to the eruption.  

\subsubsection{In situ Observations}

\begin{figure}  
	\centerline{
	\includegraphics[width=0.95\textwidth,clip=]{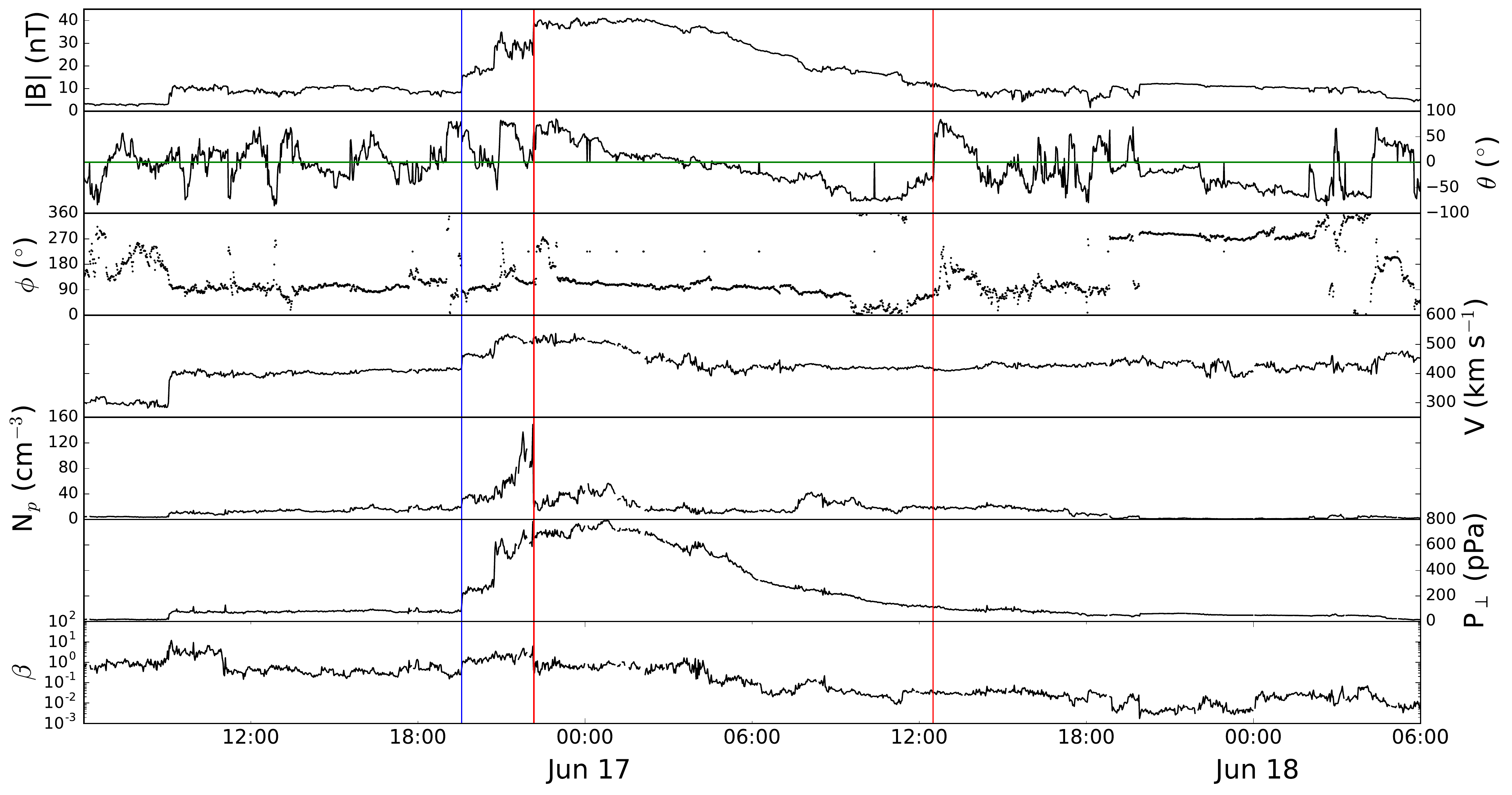}}
	\centerline{
	\includegraphics[width=0.45\textwidth,clip=]{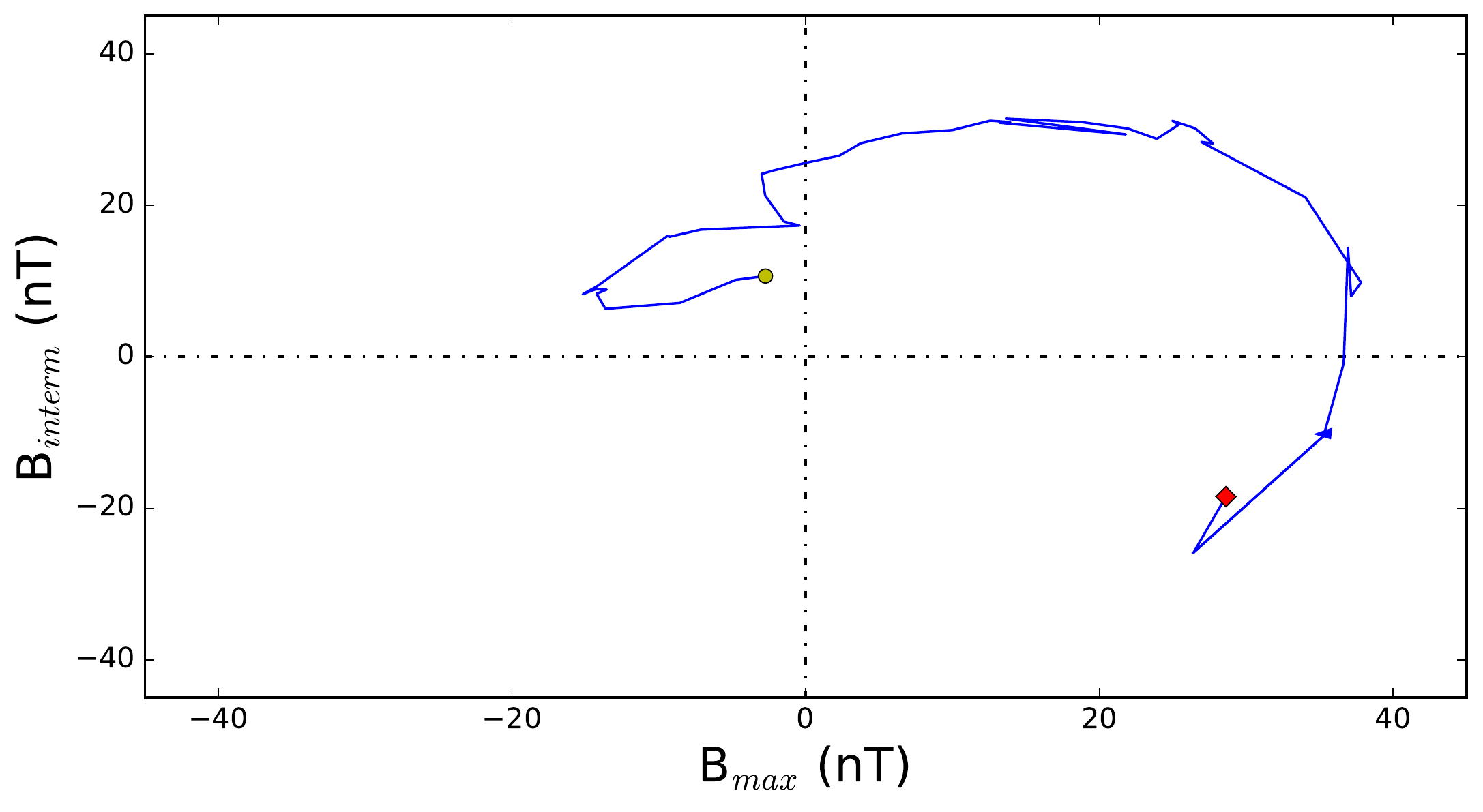}
	\includegraphics[width=0.45\textwidth,clip=]{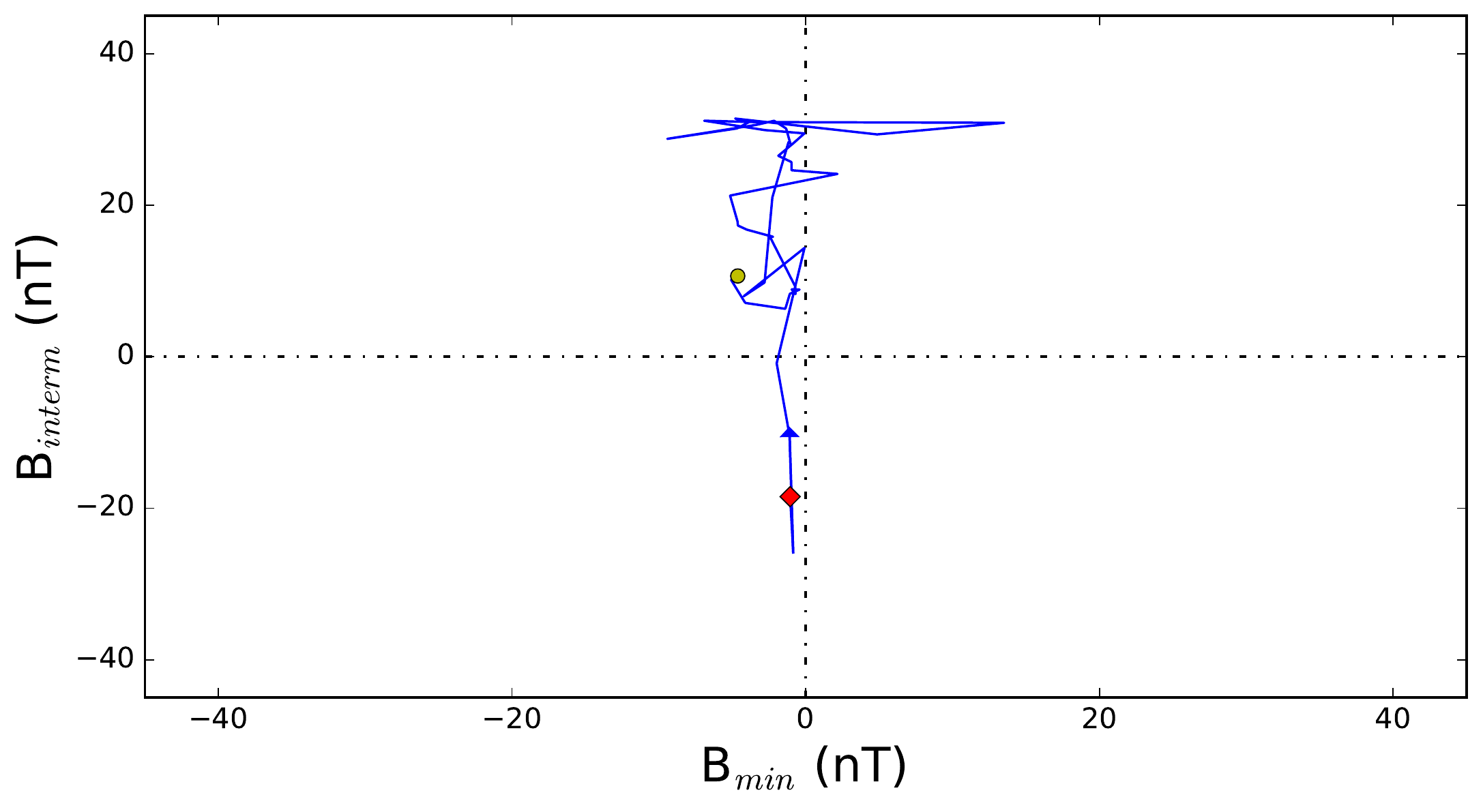}
	}
	\caption{\textit{Top:} The June 2012 CME as observed \textit{in situ} by \textit{Wind}. The blue line indicates the interplanetary shock, while the red lines indicate the leading and trailing edges of the flux rope. The parameters shown from top to bottom are: (a) magnetic field magnitude, (b) $\theta$ and (c) $\phi$ components in GSE angular coordinates, (d) solar wind speed, (e) proton density, (f) perpendicular pressure, and (g) plasma $\beta$.
\textit{Bottom:} Results of the MVA for the June 2012 CME, showing the rotation of the magnetic field vectors in the $B_{max}$-$B_{interm}$ plane (\textit{left}) and in the $B_{min}$-$B_{interm}$ plane (\textit{right}). The start of the rotation is indicated by the red diamond, the direction of the rotation by the arrow, and the end point by the yellow dot. The magnetic field data have been interpolated to a 20-minute cadence.}
	\label{2012_insitu}
\end{figure}

The interplanetary shock associated with the CME was detected by \textit{Wind} on 16 June 2012 at 19:35 UT. Clear flux rope signatures could be identified from 16 June at $\sim$22:00 UT to 17 June at $\sim$12:30 UT (Figure \ref{2012_insitu}, top panel).

The visual inspection of the data confirms that the flux rope is indeed of the NES-type: the magnetic field rotates smoothly from north to south and at the cloud center the field points eastward.  In addition, $\Delta\theta<0$, and $0^{\circ}<\phi<180^{\circ}$, where $\theta$ and $\phi$ are again the latitudinal and longitudinal components of the magnetic field, respectively.

The results of the MVA are shown in Figure \ref{2012_insitu}, bottom panels. The ratio of the intermediate-to-minimum eigenvalues is $\lambda_{2}/\lambda_{3} = 16$, confirming the validity of the method. The rotation shown in the $B_{max}-B_{interm}$ plane is now very clear and is consistent with our visual inspection, \textit{i.e.}, the NES-type. The orientation of the axis from the MVA is $(\theta_{A},\phi_{A})=(-28^{\circ},98^{\circ})$. The tilt angle is thus almost identical to the tilt of the PIL and suggests a low inclination flux rope.

The angle between the shock normal and the radial direction is $\alpha = 12.1^{\circ}$, which means that the spacecraft encountered the ICME close to the apex. The perpendicular pressure profile (Figure \ref{2012_insitu}f) shows a clear maximum around 17 June at 01:00 UT, which suggests that the spacecraft cut right through the center of the ICME, \textit{i.e.} group 1 (see Section \ref{subs:magneticinsitu}). 

\begin{figure}
   \centerline{\hspace*{0.015\textwidth}
               \includegraphics[width=0.45\textwidth,clip=]{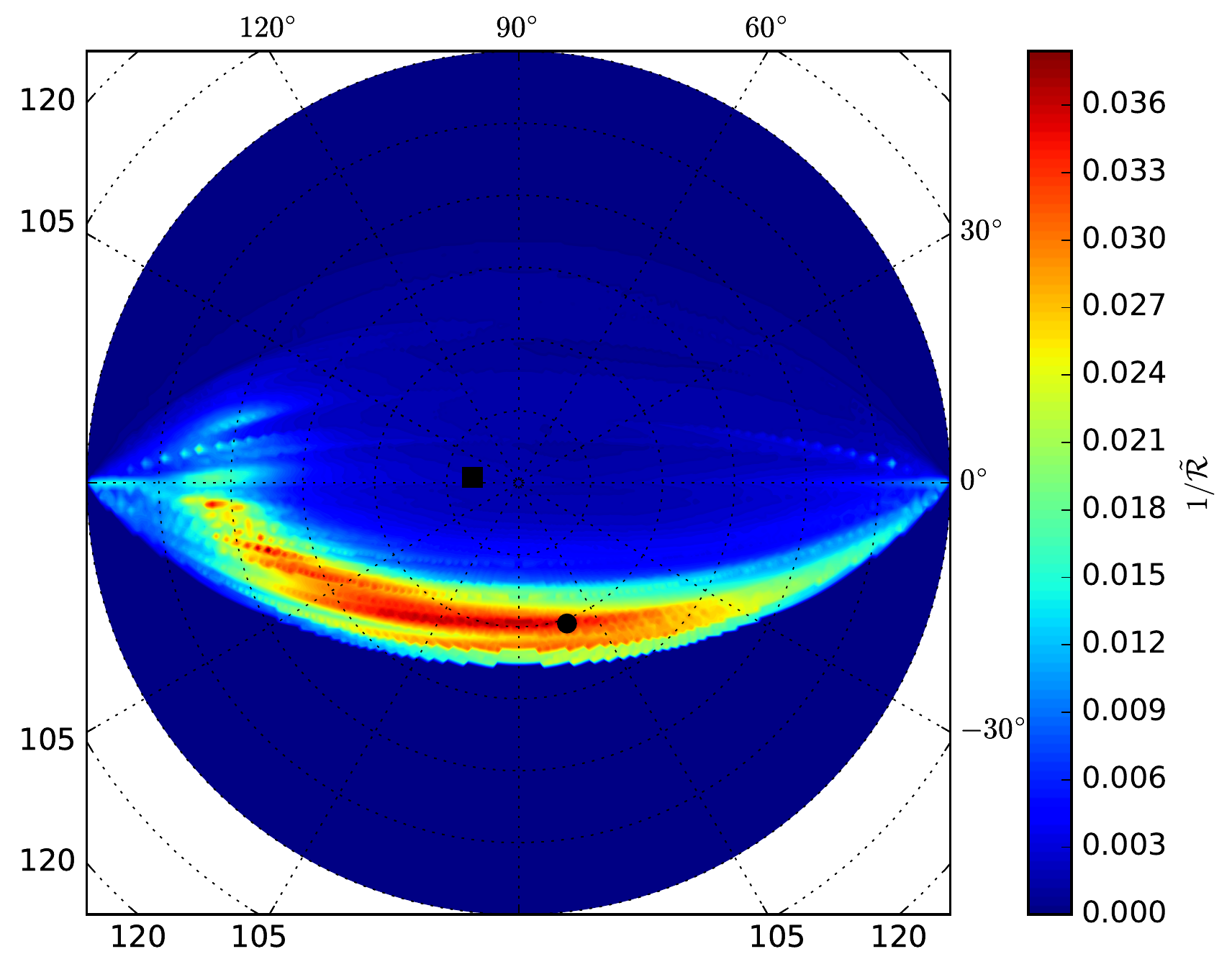}
               \includegraphics[width=0.52\textwidth,clip=]{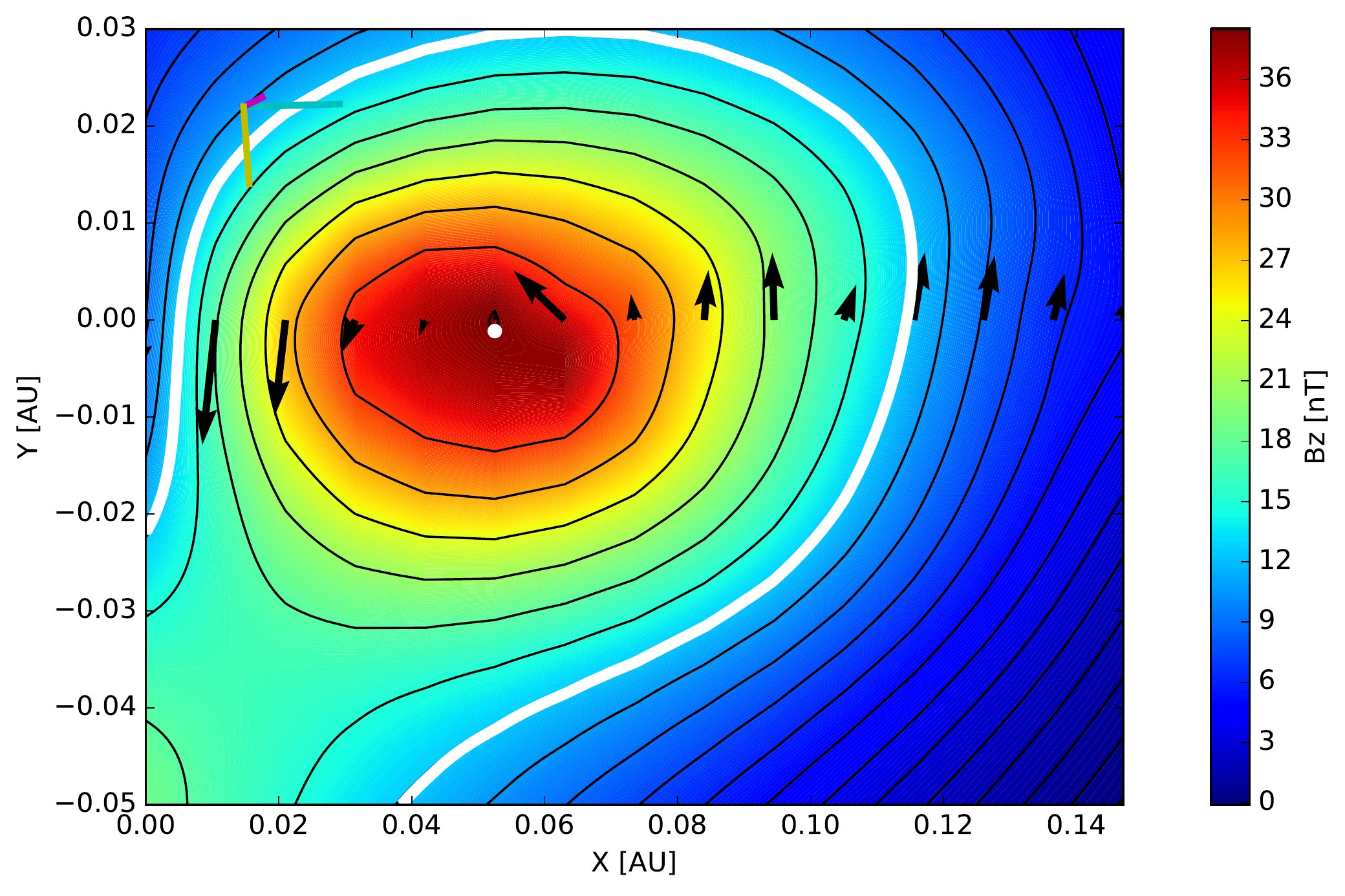}
              }
              \caption{\textit{Left:} GSR residual map for the June 2012 CME. The black dot indicates the GSR-optimized direction of the invariant axis. \textit{Right:} GSR-reconstructed magnetic field map for the June 2012 CME. The projected coordinates shown in the map are $\mathbf{\hat{x}}_{GSE}$ (cyan), $\mathbf{\hat{y}}_{GSE}$ (magenta), and $\mathbf{\hat{z}}_{GSE}$ (yellow). The Sun is assumed to be to the right of the figure.}
   \label{2012jun_gsr} 
   \end{figure}

Finally, we perform the GSR. The estimated speed of the ICME in GSE coordinates in the de Hoffmann-Teller frame is $V_{dHT} = [-462.8, -9.7, -17.0]$ km s$^{-1}$ with correlation coefficient $c=0.996$. The residual map of the event is shown in Figure \ref{2012jun_gsr}, left panel. The flux rope invariant axis has direction $\Theta = 6^{\circ}$ and $\Phi = 101^{\circ}$. The crossing distance from the flux rope nose can be estimated from the longitude of the invariant axis. If a CME is crossed near its nose, the invariant axis can be assumed to be almost perpendicular to the radial direction from the Sun. The opposite applies when the CME is crossed through one of its legs, \textit{i.e.} the invariant axis tends to be parallel to the radial solar wind flow. In this case, the longitude of the invariant axis suggests that the flux rope was crossed fairly close to its apex, consistent with our shock normal analysis above. Moreover, in agreement with the perpendicular pressure profile, the GSR suggests that the flux rope was cut centrally with a very small impact parameter.

The magnetic field map for the event is shown in Figure \ref{2012jun_gsr}, right panel. It represents the cross section of the flux rope in the plane perpendicular to the invariant axis, where the black arrows show the spacecraft \textit{in situ} observations of the magnetic field projected onto this plane. The black lines represent magnetic equipotential lines.

\section{Discussion and Conclusions}

We present a detailed analysis of the magnetic flux rope structure of two CMEs that occurred on 2 June 2011 and 14 June 2012 that were both observed in the near-Earth solar wind. We determine the axial tilt, axial field direction, and chirality of the erupting CME flux ropes using several observational proxies. 

As a first remark, it is worth noting that the eruption itself can be very complex for CMEs that appear very bright and well-structured in coronagraph images and \textit{in situ}. For example, our June 2011 event originated from between two active regions and was characterized by the double eruption of a curved filament. Only the latter eruption reached the Earth as indicated by the \textit{in situ} data. On the other hand, the June 2012 event was associated with a very well-defined flux rope \textit{in situ}, but was not associated with an erupting filament and had less distinct disk signatures. The CME appeared to originate higher up in the corona \citep[\textit{e.g.},][]{robbrecht2009, kilpua2014}. Further details about this event are discussed by \citet{james2017}.

Our analysis highlights that the presence of several alternative chirality proxies (see Section \ref{subs:chirality}) is crucial for determining the sign of the twist in flux ropes that erupt from active regions. This is because, as discussed in the Introduction, a remarkable number of active-region CMEs are not associated with filaments or the filament structure is not so distinct that it could be used to determine the sign of helicity. In addition, other techniques also have their limitations. For example, the chirality of the field in which the CME forms can be inferred from magnetic tongues only during the initial phase of emergence of an active region (as was possible for our second example, but not for our first example). Our results also show that all the proxies used to infer the chirality agree with each other. We also demonstrate the usefulness of defining the axial field direction by locating the CME footpoints in case of lack of proper filament observations. Moreover, the exact location of a filament is often difficult to determine because of projection effects. The footpoint technique can be used when clear coronal dimmings are present. Alternatively, footpoint location could be determined observationally, \textit{e.g.} from flare ribbons \citep{demoulin1996,titov1999,janvier2014}. 

However, even though both CMEs appeared non-trivial in the remote-sensing data, the flux rope types that we predicted matched the \textit{in situ} observations. For the June 2012 event, our remote-sensing prediction found a perfect correspondence with the \textit{in situ} data, while for the June 2011 event the PIL had a tilt $|\tau|=45^{\circ}$, \textit{i.e.} the dividing angle between bipolar and unipolar flux ropes. Nevertheless, the handedness of the \textit{in situ} flux rope was as expected from solar observations. The expected geomagnetic responses from SWN- and WNE-type clouds are quite different, as a WNE-cloud is expected to cause no storm, while a SWN-type cloud is expected to cause a moderate or intense storm \citep[\textit{e.g.}, see][]{huttunen2005}. However, considering an axial tilt of approximately $45^{\circ}$ implies for a right-handed flux rope an intermediate state between ``strictly'' SWN- and WNE-clouds, meaning that one expects a small amount of southward field and hence weak to moderate space weather response. 

While it is encouraging that correct estimations of the flux rope type can be obtained even for relatively complex eruptions, there are several possible ways to refine the analysis and additional parameters to take into account. For example, we have ignored the evolution of the flux rope after the eruption. To improve the current estimates, the effects of possible rotations and deflections of the flux rope in the corona and in the interplanetary space should be considered. The rotation is expected to occur mainly within the first few solar radii, where the strongest magnetic forces act on a CME \citep[\textit{e.g.},][]{gui2011,shen2011,isavnin2014}. We also emphasize the importance of determining the ``intrinsic'' flux rope type at the point of the eruption, because it is known that flux ropes of opposite chiralities tend to rotate in the opposite directions \citep{fan2003,green2007,lynch2009}. Other parameters that are important to estimate from remote-sensing observations are the magnetic flux in the flux rope and the impact parameter through the CME structure, \textit{i.e.} the crossing distance from both the CME axis and the CME nose.

%

%

%

%
\begin{acks}
EP acknowledges the Doctoral Programme in particle physics and universe sciences (PAPU) at the University of Helsinki, the Finnish Doctoral Programme in Astronomy and Space Physics, the Magnus Ehrnrooth foundation, and the Vilho, Yrj\"o and Kalle V\"ais\"al\"a Foundation for financial support. EK acknowledges UH three-year grant project 490162 and HELCATS project 400931. AJ, LG, and GV acknowledge the support of the Leverhulme Trust Research Project Grant 2014-051. LG also thanks the Royal Society for funding through their URF scheme. AI's research is supported by the European Union Seventh Framework Programme (FP7/2007-2013) under grant agreement No. 606692 (HELCATS).\newline
This research has made use of SunPy, an open-source and free community-developed solar data analysis package written in Python \citep{sunpy2015}. This paper uses data from the Heliospheric Shock Database, generated and maintained at the University of Helsinki.
\newline
\textbf{Disclosure of Potential Conflicts of Interest} The authors declare that they have no conflicts of interest.
\end{acks}

%
%
\bibliographystyle{spr-mp-sola}
\bibliography{bibliography}  
%

\end{article} 
\end{document}